\documentstyle[11pt,aaspp4,tighten]{article}

\tightenlines

\def\gsimeq
{\hbox{\raise0.5ex\hbox{$>\lower1.06ex\hbox{$\kern-1.07em{\sim}$}$}}}
\def\lsimeq
{\hbox{\raise0.5ex\hbox{$<\lower1.06ex\hbox{$\kern-1.07em{\sim}$}$}}}
\def\ion#1#2{#1$\;${\small\rm{#2}}\relax}

\begin{document}

\title{A Comprehensive Spectral and Variability Study of Narrow-Line
Seyfert 1 Galaxies Observed by ASCA: I. Observations and Time Series Analysis}

\author{Karen M. Leighly}
\affil{Columbia Astrophysics Laboratory, 550 West 120th Street, New
York, NY 10027, USA, leighly@ulisse.phys.columbia.edu}

\slugcomment{Submitted to{\it The Astrophysical Journal}}


\begin{abstract}

I present a comprehensive and uniform analysis of 25 {\it ASCA}
observations from 23 Narrow-line Seyfert 1 galaxies.  The time series
analysis is presented in this paper, Part 1, and the spectral analysis
and correlations are presented in the companion paper, Part 2.

Time series analysis shows that the excess variance from the NLS1
light curves is inversely correlated with their X-ray luminosity.
However, with a logarithmic slope of $\sim -0.3$, the dependence of
the excess variance on luminosity is flat compared with broad-line
objects and the expected value of $-1$ from simple models.  At a
particular X-ray luminosity, the excess variance is typically an order
of magnitude larger for NLS1s than for Seyfert 1 with broad optical
lines.  There is, however, a large scatter and a few objects show an
even larger excess variance.  The excess variance can be interpreted
as a time scale if the shape of the variability power spectrum, the
length of the observation and the window function are the same for all
observations, and the properties of the sample objects are shown to be
roughly consistent with this requirement.  In particular, no strong
evidence for changes during an observation in the shape or
normalization of the power spectrum was found once the systematic
errors due to the $1/f$ nature of the power spectrum was accounted for
properly.  Some of the more variable light curves are shown to be
inconsistent with a linear, Gaussian process, implying that the
process is nongaussian.  It is possible that the process is nonlinear,
but while the distinction between these possibilities is very
important for differentiating between models, such a distinction
cannot be made using these data. The enhanced excess variance
exhibited by NLS1s can be interpreted as evidence that they are
scaled-down versions of broad-line objects, having black hole masses
roughly an order of magnitude smaller and requiring an accretion rate
an order of magnitude higher.  Alternatively, NLS1s may exhibit an
inherently different type of variability, characterized by high
amplitude flares, in which case a smaller black hole mass would not be
required.

\end{abstract}


\clearpage

\section{Introduction}

Narrow-line Seyfert galaxies are identified by their optical line
properties: H$\beta$ FWHM is $<2000 \rm\,km/s$, the [\ion{O}{III}]
$\lambda 5007$ to H$\beta$ ratio is $<3$ and there are high ionization
lines and frequently strong \ion{Fe}{II} emission present in the spectrum
(Osterbrock \& Pogge\markcite{128} 1985; Goodrich\markcite{56} 1989).
Although their permitted lines may be only slightly broader than their
forbidden lines, they can be clearly distinguished from Seyfert 2
galaxies.  For example, the permitted lines are polarized differently
than the forbidden lines, indicating two emission regions (Goodrich
1989).   As shown by Boroson \& Green\markcite{14} (1992) these
emission line properties are strongly correlated; a principal
component analysis (PCA) shows that much of the variance in line
properties can be traced to a single eigenvector, despite the fact
that different emission lines are thought to originate in widely
separated parts of the AGN.  Therefore, these correlations must have
an origin in a primary intrinsic physical parameter.  The narrow-line
Seyfert 1 galaxies are located at the extreme end of the Boroson and
Green eigenvector; therefore, as a class they exemplify an extreme
value of this physical parameter.  It is important to identify this
physical parameter and to understand how it drives these correlations.

Narrow-line Seyfert 1 galaxies exhibit a few other characteristic
features.  They are seldom radio-loud (Ulvestad, Antonucci \&
Goodrich\markcite{170} 1995; Boroson \& Green\markcite{14} 1992;
Siebert et al.\markcite{159} 1998; Grupe et al.\markcite{} 1999; Moran
et al.\ in prep.).  They often have strong infrared emission (Moran,
Halpern \& Helfand\markcite{112} 1996).  Their H$\beta$ line profile
tends to have a Lorentzian rather than a Gaussian profile (Goncalves,
Veron \& Veron-Cetty\markcite{55} 1998); this may carry over to the UV
emission lines (Baldwin et al.\markcite{3} 1988; Wills et
al.\markcite{178} 1993).  There also tends to be a blue asymmetry in
the H$\beta$ line (Boroson \& Green\markcite{14} 1992; Grupe et
al.\markcite{62} 1999a).

Before {\it ROSAT}, optically or hard X-ray selected Seyfert 1
galaxies were the most intensively studied.  Narrow-line Seyfert 1
galaxies comprise less than 10\% of the AGN selected in this way (Laor
et al.\markcite{85} 1997; Piccinotti\markcite{135} 1982).  Their
paucity in such well studied samples led to the belief that NLS1s are
rare, exotic objects which should be treated as special cases and are
not representative of AGN in general.  There were hints that NLS1s may
be common in soft X-ray selected samples early on (e.g.\
Stephens\markcite{161} 1989; Puchnarewicz et al.\markcite{145} 1992;
also Halpern \& Oke\markcite{72} 1987).  The importance of NLS1s in
soft X-ray selected samples finally became clear with the {\it ROSAT}
All Sky Survey: NLS1s apparently comprise about {\it half} of the AGN
in soft X-ray selected samples (Grupe\markcite{59} 1996; Grupe et
al.\markcite{63} 1999b; Hasinger\markcite{73} 1997), and therefore,
they comprise an important subclass of AGN deserving intensive study.

Interest in studies of narrow-line Seyfert 1 galaxies intensified when
an anticorrelation between the soft X-ray photon index measured by
{\it ROSAT} and H$\beta$ FWHM was discovered (Boller, Brandt \&
Fink\markcite{12} 1996; Forster \& Halpern\markcite{47} 1996; Laor et
al.\markcite{85} 1997).  This discovery was important because it
finally provided evidence for a strong link between intrinsic
properties of the continuum emission and the Boroson
\& Green\markcite{14} 1992 eigenvector 1 through the dynamics of the broad-line
region clouds.  Several qualitative models for this correlation were
postulated.  Some models of the accretion disk predict stronger soft
X-ray emission when viewed face-on (Madau\markcite{100} 1988), and if
the broad-line region clouds orbit in a flattened distribution, a
smaller velocity width will be seen (Puchnarewicz et al.\markcite{145}
1992).  Alternatively, if the black hole mass is smaller in NLS1s,
lower Keplerian velocities will be produced.  A smaller black hole and
a higher accretion rate are predicted to produce an accretion disk
spectrum shifted into the soft X-rays (e.g.\ Ross, Fabian \&
Mineshige\markcite{150} 1992).  Recently, Wandel \&
Boller\markcite{172} (1998) have proposed a model to explain the
anticorrelation between the soft X-ray slope and H$\beta$ FWHM.  They
infer that a stronger photoionizing continuum is present in NLS1s from
extrapolation of the soft X-ray photon index into the extreme UV.  The
stronger photoionizing continuum will cause the broad-line region
clouds to form at larger radii where the Keplerian velocities are
lower.  This model does not explain the other correlations described
by Eigenvector 1, such as the anticorrelation between forbidden line
strength (e.g.\ [\ion{O}{III}] $\lambda 5007$) and \ion{Fe}{II}
emission.

A breakthrough in the understanding of NLS1s occurred with the
observation of RE~1034+39 using {\it ASCA} (Pounds, Done \&
Osborne\markcite{140} 1995).  The {\it ASCA} spectrum revealed a very strong
soft excess component dominating the spectrum below 1--2 keV, and a
very steep hard X-ray power law with photon index $\sim 2.6$.  This
spectrum is very different than the typical Seyfert 1 spectrum, which
tends to have a flatter power law with photon index typically around
1.7--1.9 and the soft excess is generally not observed in the {\it
ASCA} band pass.  This dichotomy resembles that observed between
Galactic black hole candidates (GBHC) in the hard (or low) and soft
(or high) states (e.g.\  Nowak\markcite{127} 1995), a fact that
prompted Pounds, Done \& Osborne\markcite{140} (1995) to postulate
that NLS1s are the supermassive black hole analogs of Galactic black
hole candidates in the soft state.  Soft state GBHC are thought to be
accreting at a larger fraction of the Eddington limit than hard state
GBHC; this analogy supports the idea that overall behavior
of NLS1s originates from an accretion rate that is a higher fraction
of Eddington than in Seyfert 1 galaxies with broader optical lines.

The X-ray variability properties of NLS1s are also interesting.  It
was first pointed out by Boller, Brandt \& Fink (1996) on the basis of
light curves from {\it ROSAT} PSPC observations that NLS1s frequently
display rapid short time scale variability, a result which can be
interpreted as evidence for an especially small black hole mass in
NLS1s. Based on data compiled from the literature, Forster \& Halpern
(1996) commented that very large amplitude variability on long time
scales may be a characteristic of NLS1s.  Further examples of extreme
X-ray variability from individual objects have been found from {\it
ROSAT} monitoring observations (IRAS~13324$-$3809: Boller et
al.\markcite{} 1997; PHL~1092: Brandt et al.\markcite{} 1999).
However, the X-ray variability properties of NLS1s could not be
systematically investigated using {\it ROSAT} data because of the
extremely uneven sampling.  In contrast, {\it ASCA} observations are
of similar length and they are conducted continuously, and therefore
they are amenable to systematic variability studies, although such
studies are not without difficulty due to the gaps imposed by earth
occultation and periods of high background.  It was discovered that
the excess variance from NLS1s is systematically larger than Seyfert 1
galaxies with broader optical lines.  This result is discussed in
detail here, and was first presented in
Leighly\markcite{91}\markcite{92} 1998a,b.

This paper is the first of a two part series that presents the first
uniform analysis of the X-ray variability and spectral properties from
{\it ASCA} observations of a sizable sample of Narrow-line Seyfert 1
galaxies.  This paper presents the data reduction and the time series
analysis of the {\it ASCA} data as well as a preliminary search for
spectral variability.  Part 2 presents the spectral analysis and
studies of correlations between the spectral and variability
parameters and optical emission line information from the literature
as well as from analysis of spectra in hand.

\section{Data Reduction}

\subsection{Data Selection}

All of the publically available observations of narrow-line Seyfert 1
galaxies before April 1998 were analyzed.  Proprietary observations of
PHL~1092, RX~J0439$-$45 and Mrk~142 were also analyzed.  All of the data
had undergone Rev.\ 2 reprocessing, and were reduced individually
using {\it Xselect} from the unfiltered events files.  The observation
log is given in Table 1.

{\it ASCA} analysis has by now been standardized; however, selection
criteria may be adjusted to enhance signal to noise and resolution.  A
lax set of selection criteria were used for the light curves and for
the fainter objects: minimum elevation angle above the earth's limb
(ELV) $>5^\circ$, coefficient of rigidity (COR) $>6\rm\, GeV/c$, and
in the case of the SISs, elevation above the illuminated earth
(BR\_EARTH) $>15^\circ$.  Data obtained within 4 readout periods from
the SAA were discarded, as were the data obtained within 4 readout
periods from the day-night terminator for the SISs.  Periods of
unstable attitude were also excluded.  Background monitor rates
including Sn\_PIXLm, RBM\_CONT, Gn\_H0, and Gn\_H2) were examined, and
periods when these rates were high were excluded.  SIS data taken in
Faint mode were corrected for Dark Frame Error (DFE).  The events
files were checked for telemetry saturation, but this was not a
significant problem in any of the data.  For brighter objects in which
resolution is important, stricter selection criteria were examined and
used as warranted.

\subsection{Light Curves}

 One of the goals of this paper is to compare the variability
properties of NLS1s with Seyfert 1 galaxies having broad optical
lines.  Nandra et al.\markcite{117} (1997a) presents a time series
analysis from a sample of objects chosen regardless of their optical
classification.  Their sample was dominated by broad-line Seyferts and
therefore rather than analyze a sample of broad-line objects, the
results from the broad-line objects in their sample were used for
comparison.  Thus, the comparison broad-line sample for the time
series consisted of all of their objects except the NLS1s NGC~4051 and
Mrk~335, which were reanalyzed here.  A direct comparison with the
results from Nandra et al.\ is hampered somewhat by several problems.
The largest impediment comes from the fact that the NLS1 fluxes are on
average more than a factor of 4 smaller than the fluxes of the
broad-line Seyfert galaxies considered by Nandra et al., and they also
span a wide range.  Specifically, the average count rate and standard
deviation of SIS0-only count rates from the broad-line Seyfert sample
are 1.36 and 0.72, respectively (taken from their Table 2).  In
comparison, the average and standard deviation of the count rates from
the sum of SIS0 and SIS1 of the NLS1 sample listed in Table 1 are 0.75
and 0.90.  Therefore, while the smallest bin size that Nandra et al.
consider is 32 seconds, I consider only bin sizes as small as 128
seconds.  In this case, less than 20 photons were detected in some
bins, requiring the use of maximum likelihood tests on non-background
subtracted light curves in some circumstances.  The NLS1 spectra are
generally much steeper than those from broad-line Seyfert galaxies and
therefore the count rate in the GISs was generally much smaller than
that in the SIS.  Therefore, variability tests were restricted to
combined SIS0 and SIS1 light curves.

In order to maximize the signal to noise, the upper limit on the
energy over which the light curves should be accumulated was
determined for each observation individually by examining the images
and spectra.  This was necessary because the steep spectra and wide
range in flux levels meant that for some objects, the target was not
detected to the nominal upper limit of the CCD detectors of 10~keV.
The resulting range in upper limits was 5.0 to 10.0~keV.  Conceivably
the nonuniformity in upper limits could affect the parameters derived
from the light curves, particularly if spectral variability is
present.  However, the effect is expected to be small, despite the
large range in the shapes of the deconvolved spectra. This is because
the observed counts depend not only on the source spectrum but also on
the detector effective area.  The {\it ASCA} SIS effective area peaks
around 1~keV and drops off steeply to lower and higher energies.
Therefore, combining the effective area with the steep source spectrum
means that only a small fraction of the photons are observed above
5~keV.  As an example, consider NGC~4051, a bright object which has a
relatively flat photon index among NLS1s, and which has been shown by
Guainazzi et al.\markcite{} 1996 to exhibit photon index variability.
The difference in the excess variance (see Section 3.2) from the
second observation obtained when the upper limit was 5~keV compared
with 10~keV was less than half the statistical uncertainty ($0.198 \pm
0.015$ vs $0.192 \pm 0.015$).  The lower limit in the SISs was
generally chosen to be 0.5~keV except for RX~J0439$-$45, Mrk~507,
Kaz~163, and Ark~564, where they were chosen to be somewhat higher
(0.55, 0.60, 0.60 and 0.55, respectively) in order to account for a
level discriminator applied during the observation.

To extract the light curves, the source region was enclosed in a
circle, and the background region was usually the remainder of the
chip excluding the circular region containing the source. The exceptions
were for 4-CCD mode observations when an annular background region was
used.  The source and background extraction region radii were tailored
to the flux of the observation.  Three sets were used with source
radius, background inner radius in arcminutes as follows: (3.0, 4.2),
(3.5, 4.7), (4.0, 5.3).  Subtraction of an average background rate,
rather than subtraction of the background in each time bin, improved
the behavior for very faint sources.  For the brightest objects, where
the background region is inevitably contaminated by the source
photons, the subtraction of an average background could conceivably
increase the amplitude of variability.  In practice, however, this is
a small effect, since the background is a small fraction of the source
photons; the difference in the logarithm of the excess variance was
less than 1\% for objects with SIS0+SIS1 count rates larger than
$0.5\rm \, counts\,s^{-1}$.

Figure 1 displays the light curves. For the purpose of display, a
variety of bin sizes were used, both because the large range in fluxes
among the objects means that the signal to noise for a fixed bin size
has a large range, and because a large range of luminosities were
represented and thus the variability time scales also varied widely.
Partially filled bins are also shown in Figure 1; however, for
analysis, only fully filled bins 128 seconds long were used.

\section{Variability Properties}

\subsection{Light Curves, Variability Time Scales and Spectral
Variability}

Rapid variability has been reported to be a characteristic of
narrow-line Seyfert 1 galaxies; how common is it in the {\it ASCA}
observations of NLS1s?  Following Nandra et al.\markcite{117} (1997a),
I sought variations in contiguous segments with 5 or more bins.
Fewer than 20 photons per bin were detected in some cases; thus a
maximum likelihood error analysis was used (Cash\markcite{19} 1979)
which required nonbackground-subtracted light curves.  Variability was
considered significant when the probability that the segment was
constant could be rejected with 99\% confidence.  A difference between
this analysis and that of Nandra et al. is that up to 2 bins were
allowed to be empty without the continuity being considered broken; a
two bin gap could be a result of a data dropout and would not indicate
a significant gap in the light curve from either SAA passage or earth
occultation.  The results are listed in the second column of Table 2.
One or more variable segments was found in 12 of the 23 objects, and
two or more in 6. This is comparable to the results found by Nandra et
al., although given the difference between bin size and fluxes, it is
debatable whether a direct comparison is meaningful.

Detection of variability in contiguous segments longer than 5 bins
using $\chi^2$ sensitively depends on the kind of variability (whether
a jump or a continuous increase) as well as the length of the segment.
Therefore, a complementary test of rapid variability was used.  The
$\chi^2$ was computed for contiguous strings $N$ bins long and the
number of times that significant variability is detected was counted.
The strings are not independent, so the number of detections should be
conservatively divided by $N-1$.  Results for $N=5$ are listed in
Table 2.  The variability detection rate for the two methods are
comparable, except for the second observation of NGC~4051, where more
variable segments 5 points long were discovered compared with variable
continuous strings ($161/4=40$ compared with 28).  This is probably
due to the low luminosity (therefore, rapid time scale of variability)
and brightness (therefore, good S/N) of this object.

I sought evidence for spectral variability by computing the softness
ratios as a function of time.  The softness ratio was defined as the
ratio between the count rates in the $<1.5\,\rm keV$ band to those in
the $>1.5\rm \,keV$ band.  These energy ranges were chosen to provide
good signal to noise in both the soft and hard light curves.  The
higher energy band is expected to contain essentially only power law
continuum flux, while the softer energy band may contain the power
law, soft excess, neutral absorption and absorption features from the
warm absorber.  Therefore, there is considerable ambiguity in
interpretation of observed spectral variability. It could be due to
changes in the power law index, or in the relative flux between the
power law and soft excess.  Changes in the warm absorber optical depth
are not expected to produce large changes in the softness ratio
because the of the relatively modest optical depths observed here.
Thus, this analysis is not sensitive to all forms of spectral
variability that may be discovered in a more thorough analysis beyond
the scope of this paper.

Light curves were binned by orbit for good signal to noise ratio;
orbits with exposures less than 500 seconds were excluded.  A constant
softness ratio was rejected at $>99.9$\% confidence for four objects:
NGC~4051 (in both observations), Mrk~766, IRAS~13224$-$3809 and Akn
564.  For NGC~4051, Mrk 766 and IRAS~13224$-$3809, the softness ratio
was found to be correlated with the flux using the Spearman rank
correlation coefficient at $>99$\% confidence.  In both NGC~4051 and
Mrk~766, detailed spectral analysis showed that the spectral
variability was attributable to a change in the photon index by
$\Delta \Gamma \sim 0.4$ during the observations (NGC 4051: Guainazzi
et al.\markcite{67} 1996; Mrk 766: Leighly et al.\markcite{94} 1996).
The spectral variability of IRAS~13224$-$3809 was partially
investigated by Otani\markcite{129} (1996), and appears to be more
complicated, possibly involving changes in the ratio of normalizations
of the hard and soft components and the optical depth of the $\sim 1$
keV feature.  In contrast, the softness is not correlated with the
flux in Akn~564 (Figure 2).  In three other objects a constant
softness ratio was rejected at $>98$\% confidence: NAB~0205+024,
PKS~0558$-$504, and 1H~0707$-$495.  In all of these, the flux is not
significantly correlated with the softness, except for 1H~0707$-$495,
where the positive correlation had marginal significance at 91\%.

Detection of softness variability is biased by two effects.  The
signal to noise must be high, and since spectral variability is
usually correlated with a flux change, the amplitude of variability
should be high.  Therefore, it is no surprise that significant
spectral variability was discovered in the brightest (NGC~4051,
Mrk~766, Akn~564) and the most variable (IRAS~13224$-$3809) objects.
Thus, low amplitude spectral variability would be easily detected in
bright objects, while high amplitude spectral variability would be
missed in faint objects.  For example, the $2\sigma$ softness
variability in Akn~564 in the range 1.87 to 2.43 implies an effective
$2\sigma \Delta \Gamma$ of $\sim 0.3$, which might be difficult to
measure in practice.  Thus, these two biases make it very difficult to
determine the spectral variability properties of NLS1s in general,
especially in this sample where the range in fluxes and variability
amplitude is very large.

\subsection{Excess Variance}

The excess variance is a useful parameter to characterize the
variability in unevenly sampled light curves (e.g.\ Nandra et
al.\markcite{117} 1997a).  This parameter, also known as the {\it
true} variance (Done et al.\markcite{} 1992), is found by computing
the variance of the overall light curve and then subtracting the
variance due to measurement error and normalized by dividing by the
average squared.  The square root of this parameter gives the
fractional amplitude of variability observed.

The excess variance and its error (see Nandra et al.\markcite{117}
1997a) were computed for all of the background subtracted light curves
binned at 128 seconds. The values are listed in Table 2, and they are
plotted in Figure 3 as a function of 2--10 keV luminosity, along with
the excess variances from the broad-line Seyfert 1 galaxies from
Nandra et al.\markcite{117} (1997a) as well as from NGC~7314 and
MCG--01$-$01$-$043.  This plot shows the remarkable result that the
excess variance of the NLS1s is consistently larger than that of the
broad line Seyfert 1s of the same luminosity.  This hypothesis was
tested using a two-dimensional KS test for two samples (Press et al.
\markcite{143} 1992).  Note that the KS test does not take into account
the uncertainties on the excess variances.  I find a value of the KS
statistic of 1.88 for the two distributions, with corresponding
probability of approximately 0.2\% that the two samples are drawn from
the same distribution.

Care must be taken in interpreting the excess variance and this graph.
As discussed in the appendix, the excess variance can be related to
the time scale of variability through the length of the observation.
Interpretation of this as a general property of the system requires
four assumptions: 1.) the variability power spectral shape (but not
normalization) is the same for all objects; 2.) the
light curves are stationary; 3.) the length and sampling of the time
series is the same for all objects; 4.) the intrinsically shortest
time scale of variability is not significantly smaller than the
sampling time.  The consistency of the NLS1s sample properties with
these assumptions is discussed in the Appendix A and Section 3.3.1 below.

At the low luminosity end lies NGC~4051.  It has been noted by Ptak et
al.\markcite{144} 1998 that the excess variance of this object as
shown on this plot is consistent with that of the broad-line Seyfert 1
galaxies rather than the the narrow-line Seyfert 1 galaxies.  However,
it is possible that the excess variance is low because the variability
power spectrum is different for this object compared with the others.
NGC~4051 has quite low luminosity and therefore the emission region
may be quite small; thus it may have variability power on the highest
frequencies that the other objects do not.  This will change the
excess variance by only a small amount: if the slope of the
variability power spectrum is $\alpha=1$ (where $P(f) \propto
f^{-\alpha}$), and if significant variability occurs on time scales as
short as 10 seconds, then data with better statistics will result in
only a 30\% increase in excess variance.  A more important
consideration is that NGC~4051 may have less power at low frequencies
than the other objects because of a turnover in the power spectrum.
That this may be the case can be inferred by comparing NGC~4051 with
NGC~3516 in which a low frequency turnover was discovered
corresponding to a time scale $\sim 1$ month (Edelson \& Nandra 1998).
The 2--10 keV X-ray luminosity of NGC~3516 from the {\it ASCA}
observation is $2.7\times 10^{43}\rm\,ergs\,s^{-1}$, about 70 times
that of NGC~4051.  Most time scales are predicted to scale directly
with the black hole mass; therefore the turnover time scale from
NGC~4051 should be expected at a frequency corresponding to a time
scale of less than one day. Furthermore, there is some evidence for
such a turnover perhaps in that the discrete correlation function
(DCF) for the second observation of this object decreases smoothly
toward zero at high lags.  Also, the best estimate of the power
spectral index (see Appendix A) is lower for this object than the
others.  All of these effects may conspire to give NGC~4051 an
artificially low excess variance.  Note that NGC~4051 has the lowest
luminosity in the sample by a large margin and therefore it is likely
to be the only one for which these concerns should be applicable.

The excess variance appears to be inversely correlated with the
luminosity, a result previously found for the broad line objects
(Nandra et al.\ 1997a).  To investigate this further, the regression
of the logarithm of the variance versus the logarithm of the
luminosity is computed. Note that the regression was done on the
logarithms of the luminosity and variance.  While some of the software
accommodated asymmetric error bars, most did not and the uncertainty
was taken to be the average of the error bars.  In computing the
regression, several factors are potentially important:

\begin{itemize}
\begin{enumerate} 
\item There appears to be significant scatter in the data, a fact
easily seen from the plot. Furthermore, a straight line fit to the
data is far from acceptable  ($\chi^2=1270$ for 24 degrees of
freedom (dof)).  Intrinsic scatter can be accounted for using special
methods for computing the regression (Akritas \& Bershady 1996).
\item The measured value of the variance is a function of the length
of the observation (Appendix A), and intrinsic length of the
observation will be different than the measured length if the redshift
is large.  Thus, the variance should be corrected for the redshift of
the object by multiplying by $(1+z)^{\alpha-1}$, where $\alpha$ is
the slope of the variability power spectrum.
\item The measured variance is a function of the total time spanned by the
observation, as discussed in Appendix A.  While the spans of the
observations are not widely different, there is some scatter which can
be corrected by multiplying the variance for each object by
$(T_{ave}/T_{span})^{\alpha-1}$, where $T_{ave}$ is the average amount
of time spanned by the observations (taken to be 90~ks) and $T_{span}$
is the time spanned by individual observations.
\item There is a systematic error on the excess variance which
originates in the fact that $1/f$ noise is weakly nonstationary, and
therefore, even if the shape and normalization of the variability
power spectrum doesn't change, the variance measured at two different
times will not be the same.  This fact is discussed further in Section
3.3.1 below where the size of the systematic error in the logarithm of the
variance is found to be $\sim 0.32$ and $\sim 0.46$ for $\alpha=1.5$
and 2.0, respectively.  Note that these systematic uncertainties are
quite a bit larger than the statistical uncertainties assigned to the
data.  
\item The luminosity assigned to each point is the average luminosity
during the observation; however, the luminosity has potentially very
large uncertainty due to the large amplitude variability observed in
NLS1s.  The magnitude of that uncertainty cannot be estimated;
however, the uncertainty during the observation should be at least as
large as the luminosity times the fractional amplitude of variability,
i.e.\ the square root of the excess variance.  Uncertainties in the
dependent parameter can be accounted for using the formalism developed
by Akritas \& Bershady (1996).  
\item As discussed above, the variability power spectrum from NGC~4051
may very well be turning over toward low frequencies on the time
scales relevant here.  Lying at the low luminosity end, it could
potentially bias the value of the slope.  
\end{enumerate}
\end{itemize}

It was found that a standard regression was almost certainly biased
and at least, the large scatter should be taken into account.  This,
plus whether or not NGC~4051 was included in the regression, was the
most important bias in the regression analysis; all of the other
factors resulted in changes in the uncertainty on the slope, but no
large changes on the slope itself.  Taking the redshift into account
was not very important for this sample, as only a couple of objects
(PHL~1092 and RXJ~0439$-$45) had redshifts much larger than 0.1.
Correcting the variance to a uniform span of the observation did not
produce a large difference in the regression slope; this is further
support for the idea that the spans of the observations are more or
less homogeneous.  The effect of the systematic error on the excess
variance was to drastically improve the quality of the standard
regression fit, yielding $\chi^2<47$ and as low as $17$ for 24 dof.
Including the luminosity uncertainty had little effect.  Taking all of
the effects above into account resulted in slope estimates of $-0.28
\pm 0.07$ and $-0.26 \pm 0.07$ for $\alpha=1.5$ and 2.0, respectively,
increasing to $-0.31 \pm 0.16$ and $-0.28 \pm 0.15$ when NGC 4051 was
excluded.  Considering a variety of different combinations of the
corrections listed above gave a range of slopes of $-0.26$ to $-0.31$
when NGC 4051 was included, and from $-0.30$ to $-0.37$ when it was
not, where the typical uncertainty was $0.1-0.15$.  The slope for the
broad-line objects was found to be $-0.81 \pm 0.10$ assuming intrinsic
scatter to be present.  Thus even when possible biasing factors are
taken into account, the derived regression slope is rather flat for
NLS1s compared with broad-line objects.

The presence of spectral variability can also be investigated by
measuring the excess variance in the soft and hard bands as defined in
Section 3.1.  The results (Figure 4) show that in general, the
amplitude of variability is larger in the soft X-ray band than in the
hard X-ray band.  This is likely to be an indication that the spectrum
is softer when it is bright; however, detailed analysis beyond the
scope of this paper is necessary to verify this.  This method for
investigating spectral variability is also not very robust, a fact
that is illustrated by the result from Akn~564: The excess variance in
the soft and hard bands are consistent (soft: $0.041 \pm 0.005$; hard:
$0.042 \pm 0.005$), apparently indicating that there is no spectral
variability present.  However, the softness ratio as a function of
time shown in Figure 2 illustrates that significant spectral
variability is indeed observed.

\subsection{The Structure of NLS1 Light Curves: Nonstationarity,
Nonlinearity and Nongaussianity}

Examination of the light curves in Figure 1 and comparison with those
in Nandra et al.\markcite{117} (1997a) reveal immediately the impression
that the structure of some of the NLS1 light curves is different than
those from the broad-line Seyfert 1 galaxies.  For example,
IRAS~13224$-$3809 shows large amplitude flares, both in this {\it ASCA}
observation and in {\it ROSAT} monitoring observations (see also
Boller et al.\markcite{11} 1997); such flaring is present in other
objects including especially 1H~0707$-$495 and to a lesser extent
RXJ~0439$-$495 and PHL~1092.  The fact that particular AGN can be
distinguished on the basis of the shape of its light curves is very
important for the study of X-ray variability in AGN in general,
because it means that we have a hope of understanding the origin of
X-ray variability of AGN.  

\subsubsection{Stationarity}

Ideally, the parameters obtained in time series analysis should not
depend on when the observation was done; i.e. the time series should
be {\it stationary}.  The lightcurves from AGN on time scales of a few
days are known to be characterized by a steep power-law power spectrum
($P\propto f^{-\alpha}$, where $\alpha \sim 1$ -- $\sim 2$; e.g.
Lawrence \& Papadakis 1993). Therefore, even if the parameters of the
power spectrum (the normalization and slope) are not changing, other
parameters derived from the light curve, including the mean and excess
variance. will change.  Thus, processes with such power spectra
(termed ``$1/f$'' generically) are described as being ``weakly
nonstationary'' (Press and Rybicki\markcite{} 1997) and this fact
complicates time series analysis.  However, as discussed below, the
excess variance will not change by an arbitrary amount.  Note that the
power spectrum must turn over to $\alpha<1$ at low frequencies, or the
total power would diverge, and therefore the light curves should be
stationary on long time scales.

To investigate their stationarity properties, the mean and excess
variance of the first and second halves of the light curves were
computed (Figure 5).  The average fluxes in the first and second
halves of the light curves are nearly the same.  This strong
correlation is mostly driven by the large range in fluxes in these
observations. The biggest difference between first and second half
count rates is exhibited by Mrk~766 which shows an abrupt rise in flux
during the observation accompanied by a change in the photon index
(Leighly et al.\markcite{94} 1996).

In contrast, the error bars on the excess variances indicate that this
parameter differs between the first and second halves of the
observation by a large amount for some objects.  However, this result
reflects the weakly nonstationary nature of the $1/f$ variability, and
does not mean that either the slope or the normalization of the
assumed power-law power spectrum has changed.  In other words, the
error bars on the variance are measurement errors that reflect the
sampling and noise properties of the data and they do not reflect the
differences in the variance that is expected due to the $1/f$ nature
of the variability.  The expected value of the variance can be
estimated using simulations.  One thousand long light curves with
power spectral indices 1.5 or 2.0 were generated.  A section of the
light curve was randomly chosen, then resampled and rescaled to match
the variance of the observed light curve.  Another section of the long
light curve was randomly chosen, resampled, and then rescaled using
the same scaling factor as applied to the first section.  Finally the
excess variance of the second section was measured.  The resulting
distribution of the log of the variance turned out to be approximately
Gaussian.  The 68-percentile values of the distribution clustered
around a single value for each $\alpha$, suggesting that this is an
intrinsic property related to the slope of the power spectrum.  This
makes sense, since if the normalization of the power spectrum is set
by the variance of the observed light curve, then the variance from
repeated observations of the same length should not differ by an
arbitrary amount from one another.  For $\alpha$ of 1.5 and 2 the
average and standard deviation from 22 observations (excluding the
second observation of IRAS~13349+2438, Mrk~507 and Kaz~163 because of
the low level of variability) of the log variance 68 percentiles were
$0.317\pm 0.027$ and $0.465 \pm 0.027$, respectively.  These limits
are plotted on Figure 4.  Only three objects (I~Zw~1, Mrk~766, and
PG~1244+026) clearly fall outside of the boundaries from $\alpha=2.0$.
The light curves of all three of these have structures resembling
steps or transitions to a higher state, which naturally causes a large
amplitude of variability delivered in the first half of the
observation, and relatively little in the second half.  The fact that
all of the other observations are consistent with the boundaries
indicates that on these time scales the properties of the power
spectrum do not change.

Previously, claims of nonstationarity, i.e.\ changes in the shape or
normalization of the power spectrum,  have been made based on
differing values of the excess variance from different observations of
the same object or on short strings of data within the same
observation (Nandra et al.\markcite{117} 1997a; Yaqoob et al.\ 1997;
George et al.\ 1998a; George et al.\ 1998b).  However, these claims
are suspect as they are based on the excess variance measurement
error.   As shown above, the systematic uncertainty originating in the
weak nonstationarity inherent in the $1/f$ noise can accommodate quite
large differences in excess variance without requiring a change in shape or
normalization of the power spectrum.

\subsubsection{Nongaussianity and Nonlinearity}

The light curves from some NLS1s appear to be characterized sometimes
by large amplitude flares.  This has been previously interpreted as
evidence for {\it nonlinearity} but it seems that another equally
plausible explanation is that it may be a consequence of {\it
non-Gaussianity}.  To illustrate this point, consider a generic model
as a function of time:

$$X(t)=c+\sum_{u=0}^{\infty} g_u e_{t-u}$$.

The $e$'s represent uncorrelated noise components indexed by $u$, the
$g_u$ are a sequence of constants which obey
$\sum_{u=0}^{\infty}g_u^2<\infty$ so that the variance is finite, and
$c$ is the mean of the process.  Correlations in time are introduced
by the summation.  A Gaussian light curve is one in which the $X(t)$
have a normal distribution.  If the process described above is
invertible, the $e_t$ are also Gaussian, and in fact they are a
zero-mean white noise process.  A primary and important attribute of a
Gaussian light curve $X(t)$ is that it is immediately true that a
linear model will be the best fitting one in the least squared sense.

If the $e_t$ are Gaussian, the summation will have zero mean.  Thus
the constant $c$ must be greater than zero, and in fact greater than
the summation, to describe an AGN light curve in which the flux must
always be greater than zero.  With these assumptions it can be shown
that the standard deviation $\sigma$ divided by the mean {\it \=x} from
the process must be less than 1 (Green\markcite{58} 1993).

Confusion is apparent in the literature surrounding the converse of
this idea.  What is implied if the standard deviation divided by the
mean is greater than 1?  One possibility is that the above equation
does not adequately describe the light curve and additional components
proportional to $e_{t-u}e_{t-v}$ are necessary.  Such an equation is
{\it non-linear}, since it would be quadratic in the noise term.
Because of this interpretation, several people have use the criterion
that $\sigma/${\it \=x}$>1$ to claim detection of nonlinear
variability (Green\markcite{58} 1993; Boller et al.\markcite{11} 1997).

In fact, there is another explanation.  The assumption that the $e_t$
are Gaussian may not be justified for AGN light curves.  The $e_t$ may
be non-Gaussian, and therefore, they may not have zero mean.  Then the
constant $c$ may not be necessary, and the inequality $\sigma/${\it \=x}$<1$
is no longer implied by the linear model.  The confusion here stems
from the fact while a Gaussian process implies that the model must be
linear, the converse, that a non-Gaussian process implies a non-linear
process, is not true.

That non-Gaussian noise $e_t$ may be applicable for AGN light curves
is easily imagined.  Consider a ``threshold'' process: one in which
the input signal is Gaussian, but it is only amplified if it reaches a
certain threshold.  A specific example would be electron-positron pair
creation by a thermal plasma.  Only the highest energy electrons in
the tail of the Maxwellian distribution participate in the pair
creation.  Threshold processes may produce flares of X-ray emission,
and the total signal is comprised of the sum of the flares.

Can non-Gaussianity be detected in the {\it ASCA} data from AGN?  The
answer is, yes, to an extent, but not absolutely because of the $1/f$
nature of the power spectrum.  Generally speaking, if the variability
is uncorrelated, non-Gaussianity can be established by comparing the
flux distribution with a Gaussian one.  However, if correlations are
present, as the $1/f$-type power spectrum shows that there are, this
criterion is no longer valid.  In this case, functions based on the
{\it skew} and {\it kurtosis} or the 3rd and 4th moments of the
distribution can be used (Lomnicki\markcite{98} 1961).  This method,
special requirements imposed by the peculiarities of the {\it ASCA}
data, and the development of a ``skew parameter'' are discussed in
Appendix B.

Figure 6 shows the derived skew parameter as a function of the excess
variance for the narrow-line Seyfert 1 galaxies.  Because the skew
parameter represents the number of sigma of detection of a significant
skew, lines are drawn at $-1$ and $1$ to denote the range in which the
light curves are not significantly skewed, and are indistinguishable
from Gaussian. There is a general correlation between the excess
variance and the skew.  Because of the way it is defined, significant
skew does not imply significant excess variance.  However, a light
curve with large excess variance must be significantly skewed because
the flux cannot be below zero.  All of the significant skews are
positive; this is also expected since the flux must be positive.  This
figure demonstrates the detection of nongaussian variability in
several of the more variable objects.

A Monte-Carlo ``toy'' variability model aids in the interpretation of the skew
parameter--variance result. Light curves were constructed using a
superposition of flares. It has been shown that if a light curve is
comprised of a sum of flare elements, then the power spectrum of that
light curve will have the same slope as the power spectrum of an
individual flare (e.g. Papadakis \& Lawrence 1995).  Therefore, the
basis flare element used had the form discussed by Papadakis \&
Lawrence (1995) (Equation C2).  The parameter $\beta$ was chosen equal
to 0.25 in order to produce a power spectrum with slope $\alpha=1.5$
($\alpha=2(1-\beta)$), and their parameter $c$ was arbitrarily set to 0.5.
The flares had a power law distribution of amplitudes, and the maximum
amplitude was 100 times the minimum amplitude.  The decay time scale
was scaled to the amplitude such that the shape was the same for all
flares within a simulation.  It should be noted that exponentially
decaying flares were used, but in principal the results should be the
same if exponentially increasing flares had been used instead.  Each
flare was well sampled, even into the exponential tail, with 50 points
used to describe the smallest flare.  The adjustable parameters of the
model are the slope of the amplitude distribution $\gamma$, the ratio
of the exponential decay time scale to the amplitude $Nrat$, and the
total number of flares in the simulation $Ntot$.  The total number of
flares were distributed over a grid 500000 points long, and then to
avoid minor end effects, only the central 250000 points were used.
Before analysis, the resulting light curve was rebinned by a factor of
25, yielding resulting simulated light curves 10000 points long.  From
these, the power spectrum was computed, logarithmically rebinned by a
factor of 50, and the slope measured to confirm the input $\alpha=1.5$
slope.  These showed in general a very good correspondence.  The
variance and skew of the light curves was also taken.
Eighteen sets of simulations were done, for all combinations of the
following parameters.  The total number of flares used was 2500, 7500
and 25000.  The slope of the amplitude distribution was chosen to be
$-0.5$, $-1.0$, and $-1.5$.  $Nrat$ was chosen to be 5 or 10.  The
results for the variance and skew are seen in Figure 7.  

These simulations illustrate that the excess variance can be increased
in a number of different ways.  First, the light curve can be
compressed in time.  Physically, this should correspond to the
situation where the black hole mass is smaller but the processes
producing the X-ray variability are the same.  Second, the number of
events per unit time can be reduced, or the ratio of the amplitude of
the event to the decay time scale can be reduced.  Physically, this
may imply that either the processes producing the X-ray variability
are not the same in NLS1s (i.e.\ fewer but larger bursts are
produced), or the amplitude of the events has been enhanced (i.e.
perhaps modified by beaming).  This posses an essential ambiguity that
cannot be addressed with these data.

This analysis shows that the light curves are non-Gaussian; could they
also be nonlinear?  This is a very important question, since detection
of nonlinearity can provide firm constraints on the type of models
which can 
explain the results.  If detected, it directly implies that the variability
must be at least partially coherent; that is, there must be a single
emission region, or there must be coupling between separate emission
regions.  Only models which fulfill this constraint are consistent
with nonlinear variability.  An example would be the self-organized
critical model (see e.g.\ Leighly \& O'Brien\markcite{96} 1997).

Can nonlinearity be distinguished from nongaussianity in the more
variable objects from this sample?  The answer is no, for the
following reasons.  The surrogate data method (Theiler et
al.\markcite{166} 1992; Kantz \& Schreiber\markcite{77} 1997; Leighly
\& O'Brien\markcite{96} 1997) provides a simple and sensitive way to detect weak
nonlinearity. Surrogate or fake data with the same linear properties
as the real data are constructed from the variability power spectrum
of the real data by scrambling the Fourier phases.  Then the
properties of the real and surrogate data are studied using a
nonlinear statistic; if the results differ significantly, nonlinearity
is indicated.  However, care must be taken when applying this method.
The surrogates described above are Gaussian and therefore can only be
used to test against the hypothesis that the time series is a linear
combination of Gaussian noise.  The surrogate data method must be
modified to test against the possibility that the time series is a
linear combination of non-Gaussian components.  This is done by
adjusting the distribution of amplitudes of the surrogates to match
that of the real data, a modification of the original procedure that
would be appropriate for these data. A second, more severe problem is
that the surrogate data method is sensitive also to nonstationary in
the light curves.  Therefore, a false detection of nonlinearity can be
obtained merely if the mean and variance are changing with time (e.g.\
Timmer\markcite{167} 1998). This method was used to search for
nonlinear variability in the broad-line radio galaxy 3C 390.3 (Leighly
\& O'Brien\markcite{96} 1997).  Gaussian surrogates were used when
amplituded adjusted surrogates would have been appropriate;
furthermore, the 3C~390.3 light curve is clearly nonstationary.
Nevertheless, the fact that quiescent periods occur before and after
flares, which is evidence for coherent structure in the light curves,
still supports nonlinearity in the 3C~390.3 light curve.

The surrogate data method cannot be applied directly to the NLS1 light
curves because of the gaps.  However, a modification of the surrogate
data method which conserves the Lomb periodogram (Schmitz \&
Schreiber\markcite{153} 1998), could be applied conceivably.  However,
the same caveat as above holds true: the method will be highly
sensitive to the nonstationarity as well as the nonlinearity, and
therefore much longer observations are needed to make a serious
attempt.  Preliminary tests on the second observation of NGC~4051
indicate that the surrogates are not significantly different from the
real data; however, there are currently some difficulties in defining
a good test statistic (Schmitz, P. comm. 1998). Finally, there is no
evidence suggestive of coherence, as was found in 3C~390.3; however,
poor signal to noise and interrupted coverage could be hampering
recognition of coherence signatures.

The toy model presented above shows that large excess variance can be
produced by a non-gaussian model.  Can we also expect large excess
variance if the model is nonlinear?  This problem has been partially
investigated already by Paltani \& Courvoisier (1997).  They construct
a ``non-Poisson process'' model which may be somewhat similar to a
one-dimensional SOC model.  Rather than assuming that events occur
randomly in time, they modify the usual Poisson probability to reduce
the mean time between events when an event appears.  This results in
enhanced variability as a function of luminosity, simply demonstrating
that it is possible that a nonlinear SOC model may produce enhanced
variability.  Vio et al.\ 1992 also discuss high amplitude variability
in the context of nonlinear models.

It is important to stress that neither non-Gaussianity nor
nonlinearity can be recognized from the power spectrum alone.  Many
processes can produce a steep power spectrum, including self-organized
critical models (e.g.\ Mineshige, Takekuchi \& Nishimori
\markcite{110} 1994) or model in which the emission from hot spots on
the accretion disk is modulated by Doppler motion and gravitational
lensing (e.g.\  Bao \& Abramowicz\markcite{4} 1996 and references
therein).  Begelman \& DeKool\markcite{5} 1990 discuss other models
that can produce a steep variability power spectrum.  The important
point to note is that some of these models are nongaussian and others
are nonlinear and the slope of the power spectrum cannot be used to
differentiate between these types of models.  The limitations of the
power spectrum in differentiating between types of variability is also
discussed by Press (1978). 

In summary, using a parameter based on the skew of the flux
distribution, and a series of simulations, it was found that linear,
Gaussian variability could be ruled out in 7 and 2 objects with
significance greater than 1 and 2 sigma, respectively.  This result is
interpreted as evidence of a linear non-Gaussian process, at least; a
nonlinear process is also a possibility.  These two possibilities
cannot be distinguished using these data.

\section{Discussion}

\subsection{Variance vs luminosity relationship}

Barr \& Mushotzky (1986) first discovered that the variability time
scale is correlated with the luminosity.  They used a ``doubling time
scale'', defined as the time required for the flux to change by a
factor of two, and derived by extrapolating observed variability.
This method should not produce very reliable results when the
variability power spectrum is proportional to $f^{-\alpha}$.
Papadakis \& Lawrence (1993) examined long, uninterrupted observations
of AGN made with {\it EXOSAT}.  They also found that the variability
time scale is correlated with the luminosity, basing the variability
time scale on the observed amplitude of the power spectrum at a
particular frequency.  Green, McHardy \& Lehto (1993) use a similar
method on the same data and find a similar result.

In this paper (also Nandra et al.\ 1997a) the variability was studied
using the excess variance.  How can that be interpreted as a
variability time scale?  The dependence of the excess variance on the
length of the observation is discussed in Appendix A (also Press \&
Rybicki\markcite{142} 1997).  The observed variance is proportional to
the length of the observation $T_{obs}$ as $T_{obs}^{\alpha-1}$ where
$\alpha$ is the slope of the power spectrum ($F(f) \propto
f^{-\alpha}$) \footnote{Note that Nandra et al. 1997a neglected the
dependence on the slope of the power spectrum in their discussion.}.
The length of the observation can be related to the time scale of
variability through the scale-invariant nature of the assumed power
law form of the variability power spectrum. If the time scale of
interest $T_i$ is defined as the time that it takes the observed
variance to reach a particular value, then the observed variance is $
\propto T_i^{1-\alpha}$ (see also Lawrence
\& Papadakis 1993).

Simple, familiar arguments link the time scale to physical parameters.
A lower limit on the time scale of variability is the light crossing
time $\Delta t=R/c$, where $R$ is the size of the emission region.
The X-ray emission region may have radius about $10 R_G$, where $R_G$
is the gravitational radius ($R_G=GM/c^2$) and $M$ is the black hole
mass.  The black hole mass can be related to the Eddington luminosity
by $M=L_{Edd}/1.25\times 10^{38}$ which in turn can be related to the
observed luminosity through the accretion rate by $L=L_{edd} \eta \dot
m c^2$ where $\eta$ is the efficiency of conversion of accretion
energy to radiation and $\dot m$ is the specific accretion rate in
units of the Eddington rate.  The result is that $$variance \propto
(\frac{L}{\eta \dot m})^{1-\alpha}.$$

Figure 8 displays lines determined by this formula, separated by a
decade of the variable parameter (either $\eta$, $\dot m$ or the
produce) and superimposed on Figure 3. Two cases, $\alpha=1.5$ and
$\alpha=2$, are shown.  For a particular hard X-ray luminosity, the
excess variance is about an order of magnitude larger for the NLS1
compared with the Seyfert 1 galaxies with broad optical lines.  This
formula shows that, at a given luminosity, if $\alpha=2$, then either
the accretion rate or the efficiency must be an order of magnitude
larger for NLS1s.  That factor increases to two orders of magnitude
for $\alpha=1.5$, because of the shallower dependence that excess
variance has on the time scale.  While most of the NLS1s have excess
variance about one order of magnitude larger than the broad-line
objects, there are a few objects which show much greater excess
variance.  Specifically, PHL~1092 has 2--3 orders of magnitude higher
excess variance than would be predicted for a broad-line Seyfert 1
galaxy.  This would be interpreted as 2--3 orders of magnitude larger
variable parameter assuming $\alpha=2$, increasing to 4--6 orders of
magnitude assuming $\alpha=1.5$.

The superimposed lines on Figure 8 show that the trend of the log of
the variance versus the log of the luminosity is very nearly precisely 
proportional for the broad line objects (i.e. $var \propto L^{-1}$).
This is similar to the previous result found by Lawrence
\& Papadakis (1993) that $T \propto L^{0.96}$.   
However, as discussed in Section 3.2, even when various factors which
may bias the result for the NLS1s are taken into account, the
relationship is quite a bit flatter ($var \propto L^{\sim-0.3}$).

The case where $T_i \propto L$ can be simply explained in two
different ways.  First, the black hole mass may be larger for more
luminous objects, corresponding to a larger emission region and longer
time scales of variability.  This scenario produces a higher
luminosity (because of the larger luminosity of individual events) and
the variability is reduced because the time scales are proportionally
longer.  Alternatively, a Poisson variability model has been proposed
in which the AGN X-ray emission region consists of a large number $N$
of small fluctuating emission regions, each of which emits a fixed
luminosity (e.g. Paltani \& Courvoisier 1997; Green, McHardy \& Lehto
1993).  Then, more luminous objects are simply characterized by a
larger number of emission regions.  If the number of emitting regions
at one time follows a Poisson distribution, the excess variance would
be proportional to $1/N$.  It is necessary to measure a physical
system time scale (distinguished from $T_i$) to differentiate between
these two models, and it is not obviously possible to do this with the
data presented here.  However, the presence of apparently coherent
variability on long time scales from the rather luminous AGN 3C~390.3
(Leighly \& O'Brien 1997) suggests that the size scale is indeed
larger in more luminous objects.

For the NLS1s, the origin of the flat variance versus luminosity
relationship is not known.  The flat slope would appear to rule out
the Poisson variability model, or a model in which both the time scale
and luminosity of events scales simply with black hole mass.  First,
however, it is possible that the NLS1s do not represent a homogeneous
class of objects with respect to their variability characteristics.
In this case, a regression would certainly give misleading results.
Paltani \& Courvoisier (1997) find that AGN UV variability on long
time scales also has a rather flat dependence on luminosity ($\sigma
\propto L^{-0.08}$, equivalent to $var \propto L^{-0.16}$).  They
explore several mechanisms which would produce such a dependence.
First, they consider the idea that a more luminous object would
produce more luminous single events, without the rate of events being
reduced.  This would increase the luminosity without changing the
variability characteristics and thus it would tend to produce a flat
dependence of variance on luminosity.  Another scenario retains the
same flux per event independent of the luminosity, but postulates that
the process is not Poissonian.  In other words, events don't occur
randomly, but rather there is an enhanced probability for another
event to occur if one has recently occurred.  This model seems to be
similar to a simplified SOC model.  They find that a flattened variance
versus luminosity relationship can be attained over a limited range of
luminosity such that the rate of events is approximately one per time
bin.  Since it seems intuitive that the luminosity, time scale and
amplitude should scale together roughly with black hole mass, neither
of these mechanisms appear to have an obvious or immediate physical
motivation.

\subsection{Models for Enhanced Excess Variance in NLS1s}

The previous section showed using simple scaling arguments that the excess
variance can be related to the luminosity through the accretion rate
$\dot m$ and the efficiency of conversion of accretion energy to
radiation.  An important assumption inherent in this analysis is that
the hard X-ray luminosity is characteristic of the absolute accretion
rate times efficiency, $\eta\dot M$.    This assumption may be
questioned; for example, since NLS1s have overall steeper spectra,
more of the radiation may be coming out in the soft X-rays.  If that
were the case, however, the difference between NLS1s and broad line
objects would be exacerbated, as the luminosity for the NLS1s would be
underestimated.  

Before discussing physical models, it is worth briefly reviewing
relevant time scales.  Detectable variability was observed from the
light curves of nearly all of the NLS1s considered here.  The
variability time scales observed provide basic constraints on the
mechanism producing the variability.  Standard accretion disks have a
series of natural time scales.  The shortest time scale of variability
is the dynamical time scale, or Keplerian period.  At the last stable
orbit for Schwarzschild geometry, $r=3R_S$ where $R_S$ is the
Schwarzschild radius, and the dynamical time scale $t_{dyn}=454 M_6$
seconds, where $M_6$ is the black hole mass in units of $10^6
M_\odot$.  The thermal time scale $t_{th}$, or the time to equilibrate
an energy disturbance in the disk, is longer than the dynamical time
scale: $t_{th}=\alpha^{-1}t_{dyn}$, where $\alpha$ ($<1$) is the disk
viscosity parameter. The viscous or radial drift time scale, or the
time for accretion perturbations to propagate, is longer still:
$t_{vis}=${\cal M}$^2 t_{th}$.  {\cal M} is the Mach number in the
disk, {\cal M}$=v_{dyn}/c_s$ where $c_s$ is the sound speed, $c_s^2
\approx P/\rho$ and $P$ and $\rho$ are the pressure and density
respectively.  For an $\alpha$ disk and a $10^6 M_\odot$ black hole,
under the best possible circumstances, the viscous time scale is about
half a year, and therefore perturbations on this time scale are
unlikely to be driving the rapid X-ray variability we see.  The
thermal or dynamical time scales are more likely candidates to drive
the variability.  Alternatively, it is possible that the variability
may originate in the hard X-ray component and then be reprocessed in
the soft X-ray component; however, because the photon index is steeper
than 2, it is energetically impossible for the hard X-rays to
completely power the soft X-rays by reprocessing and therefore the
applicability of this process to NLS1s is limited.  However, for
proposed dynamo models for this process (e.g. Galeev, Rosner \&
Vaiana\markcite{48} 1979; Haardt, Maraschi
\& Ghisellini\markcite{70} 1994; Di~Matteo\markcite{31} 1998), the
variability time scale should be similar to the dynamical time scale.

What is the origin of the enhanced excess variance in NLS1s?  A simple
explanation that has been proposed previously is that NLS1s are
characterized by an enhanced accretion rate relative to their black
hole mass; i.e. $\dot m$ is larger.  The simplest implication of this
explanation would be that assuming that the efficiency of conversion
of gravitational potential energy to radiation is the same in all
objects, then NLS1s have relatively smaller black hole masses for a
given X-ray luminosity and since the size scale is therefore smaller,
the variability can be more rapid.  Thus, the structure of a light
curve from an NLS1 may be just the same as that from a broad-line
Seyfert galaxies, but compressed to $\sim 1/10$ of the length.
Because the measured excess variance is a function of the length of
the observation, a higher excess variance is measured from the NLS1s.
This hypothesis is supported by the shape of the X-ray spectrum which
for some objects resembles that of Galactic black hole candidates in
the high state (e.g. Pounds, Done \& Osborne 1996).

However, this is not the only explanation.  It is instead possible
that the structure of the light curve from NLS1s is different, and the
amplitude of the variability is enhanced.  This can be viewed as an
enhancement of $\eta$ rather than $\dot m$ in the equation, but
physically it could imply that the mechanism of the variability is
different in some way in NLS1s compared with broad-line objects.  Note
that this view may also be consistent with an enhanced accretion rate, if
the effect of the enhanced accretion rate is that accretion solution
and the mechanism for variability is different.  This view is
supported by the results of the toy model simulations presented in
Section 3.2.  These show that the same values of excess variance can be
obtained by models consisting either of many very narrow flares or few
isolated large flares.

The structure of the light curves was examined quantitatively in
Section 3.3 and Appendix B.  It has been shown that if correlations are
present in the light curves, functions based on the skew and kurtosis
can be used to detect non-Gaussianity (Lomnicki\markcite{98} 1961).
While these functions cannot be applied directly here because of the
$1/f$ nature of the power spectrum, the non-Gaussianity can still be
examined using a Monte-Carlo approach by comparing the results from
the data with those from many simulated Gaussian light curves.  It was
found that a few of the more variable light curves show significant
evidence that they are consistent with being non-Gaussian. The
discovery that the light curves are inconsistent with a linear,
Gaussian model is perhaps not surprising.  If the variability were
linear and Gaussian, then an additional source of constant emission
would be necessary, since the flux cannot be less than zero.  The
source of such constant emission might be difficult to explain.  No
test for nonlinearity could be applied to the light curves; the first
problem is the presence of gaps in the light curves and the second is
that as results are likely to be ambiguous due to difficulties with
nonstationarity and an insufficient amount of data.  However,
nonlinear models are likely also to produce enhanced excess variance
and skewed flux distributions; this has been partially demonstrated in
a simple non-Poisson model constructed by Paltani \& Courvoisier (1997).
Thus the lightcurves which are shown to be inconsistent with linear
Gaussian variability may be nongaussian or they may nonlinear.  This
distinction is an essential one for distinguishing between candidate
variability models.  Unfortunately it is a distinction that can not be
made with these data.

Both nongaussian and nonlinear models have been proposed for AGN
variability.  An example of a nongaussian model is the one proposed by
Abramowicz and collaborators, in which AGN X-ray emission is confined
to discrete regions (spots) on an inclined accretion disk (e.g. Bao \&
Abramowicz\markcite{4} 1996 and references therein).  Emission from
each spot is transient, and the transient emission is further modified
by Doppler effect due to their Keplerian orbits and gravitational
effects due to proximity to the black hole.  Because the spots are
independent, this is a nongaussian model. 

This non-Gaussian model has been applied to NLS1s specifically by
Boller et al.\ (1997). They propose to explain the large amplitude rapid
variability observed in NSL1s by relativistic effects associated with
emission at the inner edge of the accretion disk orbiting the black
hole.  At the inner edge of the accretion disk, motions will be
relativistic and Doppler boosting and gravitational lensing of the
emission will strongly enhance the amplitude of the variability.  The
variability will be amplified more when the spectrum is steeper and
when the inclination is larger compared with the normal.  The
dependence on the slope is not sufficient to explain the factor of 10
enhanced variability observed in NLS1s and therefore a systematically
larger inclination and/or smaller inner radius would also be required.
Also, note that the lack of periodic variability requires the soft
X-ray emitting inhomogeneities be short-lived compared with the
dynamical time scale. Note the hard X-rays also show large amplitude
variability on a similar time scale during the {\it ASCA} observation;
therefore, the hard X-rays should also be emitted at the inner edge of
the accretion disk.  Although promising, this model is not without
problems.  It may be difficult to reconcile a model based on large
inclination for NLS1s with their other properties however, since it
would require that the optical emission line clouds be moving along
along the symmetry axis, rather than in the plane of the accretion
disk, to avoid broadening the line profile.  Such a model could in
principle be tested using the iron K$\alpha$ line, since a significant
inclination Schwarzschild geometry predits observation of a
significant blue wing in addition to a blueshifted horn (Fabian et
al.\markcite{41} 1989; Matt et al.\markcite{106} 1992).

An example of a nonlinear model are the family of self-organized
critical (SOC) models.  SOC models have been applied to accretion
disks for the Galactic black hole candidate case by Mineshige and
collaborators (Mineshige, Ouchi \& Nishimori\markcite{} 1994).  For
the accretion disk case, the disk is imagined to be comprised of
numerous reservoirs which are coupled to their neighbors.  Accretion
occurs, and once a reservoir reaches a critical density, an accretion
avalanche occurs.  As the neighboring reservoirs are coupled, an
avalanche in one reservoir may stimulate avalanches in a few or many
neighboring reservoirs.  It is this coupling that makes the model
nonlinear.

X-ray emission from AGN may be too energetic to arise from the
accretion disk; however, SOC models are quite general and may be
applied in other circumstances.  SOC models were first applied to the
problem of X-ray variability from the Solar corona by Lu \& Hamilton
(1991) and since then these models have been extensively explored.
There may be therefore a parallel in the accretion disk corona.  SOC
models have been applied to accretion disk atmospheres of cataclysmic
variables (Yonehara, Mineshige \& Welsh 1997).  An avalanche model for
magnetic loops that is somewhat related to the SOC model has been
applied to Galactic black hole candidates (Poutanen \& Fabian 1999).
Alternatively, the emission region may be the base of a slow jet
(e.g.\ Mannheim, Schulte \& Rachen 1995).

Recently further support has come for the idea that the amplitude of
variability is enhanced in NLS1s, rather than the time scale of
variability being reduced.  Brandt et al.\ 1999 found rapid
variability by a factor of 4 in less than 3.6~ks in a set of {\it
ROSAT} monitoring observations of the luminous narrow-line quasar
PHL~1092.  Such high amplitude variability almost necessarily requires
relativistic boosting or beaming of the emission.

While relativistic boosting at the inner edge of an accretion disk is
a somewhat attractive model to explain the high amplitude variability
from NLS1s, it is certainly not the only site of boosting imaginable.
NLS1s are in general radio quiet, but recently several examples of
radio-loud NLS1s have been discovered (e.g. Siebert et al.\ 1999; Grupe
et al. 1999b).  These ratio of the radio to optical emission is not
large in these objects and therefore they can be considered radio
moderate.  It has been proposed that such radio-moderate or
radio-intermediate objects are have a weak compact jet and appear
radio-moderate because they happen to be viewed nearly pole-on
(Falcke, Sherwood \& Patnaik 1996).  The weak jet may also be a site
of X-rays (e.g.\ Mannheim, Schulte \& Rachen 1995), although it may be
hard to explain why the radio collimation angle should be smaller than
the X-ray, as implied by population statistics.  An additional
favorable point for this scenario is that an isolated emission region
such as a frustrated jet may behave in a coherent fashion and
therefore the variability may be nonlinear.

NLS1s are known to have softer X-ray spectra than Seyfert 1 galaxies
with broader optical lines (e.g. Part 2).  Previously, three workers
have noticed that objects with steeper X-ray spectra have enhanced
variability.  Green, McHardy \& Lehto (1993) found this trend in the
{\it EXOSAT} ME data.  They attribute this behavior to reprocessing:
Compton reflection will tend to flatten X-ray spectra and if a
significant amount of the flux is being reflected, and the reflecting
surface is sufficiently large or far from the X-ray source, the
variability will be reduced.  This model is shown to be untenable by
the data presented here, as the {\it ASCA} response is sufficiently
soft that reflection will not contribute significantly.  K\"onig,
Staubert \& Wilms (1997) showed that the {\it EXOSAT} light curves
could be described by a first order autoregressive model, and the
derived decay time scales are inversely correlated with the power law
slope; that is, steeper objects have shorter decay time scales.  They
attribute this effect to Comptonization, proposing that the X-rays
forming the harder spectra are scattered more times, resulting in a
longer time scale of variability.  This view would be supported if it
were necessarily true that the plasma temperature were the same in
NLS1s and broad-line objects.  However, the current paradigm for the
hard X-ray spectrum from NLS1s holds that, in analogy with thermal
models for black hole candidates in the high state, the plasma
temperature is lower for NLS1s and therefore their steep spectrum
arises from the same number of scatterings but smaller energy boost
per scattering (e.g.\ Pounds, Done \& Osborne 1995).  Furthermore,
observations to high energies show that the broad band X-ray to
Gamma-ray spectrum is best modeled with an optical depth of about 1
for broad-line objects (e.g.\ Zdziarski 1999).  Therefore, to produce
significant dilution in the variability, the scattering plasma must be
very much larger in broad-line objects than in NLS1s.  In this case, a
lag of the hard X-ray emission over the soft X-ray emission would also be
expected; such a lag has never been securely demonstrated in AGN.  It
is worth noting that K\"onig, Staubert \& Wilms (1997) claim that the
decay time scale is not related to luminosity; however, their sample
is very heterogeneous, including Seyfert 2 galaxies and blazars, and
if only radio-quiet objects are considered, the usual correlation is
obtained.  Finally, Fiore et al.\ 1998a discuss {\it ROSAT} HRI
monitoring observations of three quasars with flat X-ray spectra and
three with steep X-ray spectra.  They also find that the variability
is enhanced in the objects with steep spectra, and attribute the
effect to a higher $L/L_{Edd}$ in the steep-spectrum objects.

\section{Summary and Future Observations}

I present a comprehensive and uniform time series analysis of 25 {\it
ASCA} observations from 23 Narrow-line Seyfert 1 galaxies.  The
results of spectral analysis and correlations with optical emission
line properties follow in Part 2.  The primary results of this paper
are the following:

\begin{itemize}
\item The excess variance was found to be inversely correlated with
the luminosity for the NLS1s, a result which has been previously found
from Seyfert 1 galaxies with broad optical lines. However, the
dependence of variance on luminosity is somewhat shallower for NLS1s
in this sample (logarithmic slope $\sim -0.3$) compared with
broad-line Seyferts, a result that is difficult to explain with simple
models.  The excess variance is typically an order of magnitude larger
for NLS1s compared with Seyfert 1s with broad optical lines and
similar hard X-ray luminosity.  Since the sampling and length of the
observations are approximately the same, and since the variability
power spectral index appears to be consistent among the more variable
objects and constant during each observation, the excess variance can
be interpreted as a time scale of variability.
\item The structure of the light curves was examined using a parameter
related to the skew of the flux distribution, and the light curves
from several of the more variable objects were shown to be
inconsistent with a linear, Gaussian process.  This result implies
that the process is either nonlinear or nongaussian, but these two
possibilities cannot be distinguished using these data.  
\end{itemize}

Assuming that the structure of the light curves from NLS1s is the same
as those from Seyfert 1 galaxies with broad optical lines, but
compressed in time, the enhanced excess variance for NLS1s can be
interpreted as evidence that the black hole mass from these objects is
typically an order of magnitude smaller than for Seyfert 1 galaxies
with broad optical lines.  However, the discovery that the light
curves from some of the more variable objects is inconsistent with
a linear, Gaussian process suggests that the structure of the light
curves may be different for NLS1s, and characterized by high amplitude
variability.  In this case, a smaller black hole mass may not be a
requirement and beaming of the emission is a possibility.

The {\it ASCA} observations and results presented here lead to further
questions which may be answerable in the future.

\begin{itemize}
\item The problem of distinguishing between nonlinear and nongaussian
variability is very important, since it would strongly constrain the
geometry of the emission region.  This is a difficult problem in
general, but in the case of AGN light curves it is particularly
difficult because of the $1/f$ nature of the power spectrum.  Possibly
the best method currently available to detect weak nonlinearity is the
surrogate data method (Theiler et al.\ 1992); however, nonstationarity
is difficult to distinguish from  nonlinearity using that method
(e.g.\ Timmer\markcite{167} 1998), and therefore very long light
curves are necessary.  However, employing various techniques to
examine the three point correlations in evenly sampled data and
comparison with simulations is likely to yield interesting results,
and at least it may allow us to determine whether the structure of NLS1 light
curves is the same or different from those  of broad-line objects.
\item The problem of determining whether NLS1s in fact have a smaller
black hole mass than Seyfert 1 galaxies with broad optical lines can
be approached by measuring and comparing characteristic physical time
scales.  Such time scales can be obtained for example from the low and
high frequency turnovers of the power spectrum.  The low frequency
turnover requires long term monitoring, as is currently being
performed by {\it RXTE} for a number of objects, and the high
frequency turnover requires high throughput, such as will be available
from {\it XMM}, so that high signal-to-noise can be obtained on short
time scales.  Autoregressive modeling may also yield useful time
scales.  However, comparison of these time scale is only valid if the
processes producing the variability are the same, and therefore a
search for variability time scales must be combined with studies of
the structure of the light curve.  Measurement of physical variability
time scales will also be useful to determine whether the time scales
are longer in high luminosity objects.
\end{itemize}

\acknowledgements

KML gratefully thanks all those people who built and operate {\it
ASCA}.  This project could have never attained this form without the
help and support of Jules Halpern.  Many thanks go to Dirk Grupe for
many various kinds of help and advice.  Useful discussions with Karl
Forster, Toshihiro Kawaguchi, Randall LaViolette, Herman Marshall,
Francesco Paparella, and Tahir Yaqoob are acknowledged.  The following
are thanked for a critical reading of a draft: Joachim Siebert, Jules
Halpern \& Tahir Yaqoob.  KML also gratefully thanks the organizers of
the 1998 TISEAN workshop for allowing her to participate and
acknowledges useful discussions with the participants, especially
James Theiler, Holgar Kantz, Thomas Schreiber \& Andreas Schmitz.
This research has made use of the NASA/IPAC extragalactic database
(NED) which is operated by the Jet Propulsion Laboratory, Caltech,
under contract with the National Aeronautics and Space Administration.
This research has made use of the {\it ASCA} IDL Analysis System
developed by Tahir Yaqoob, Peter Serlemitsos and Andy Ptak.  This
research has made use of data obtained through the High Energy
Astrophysics Science Archive Research Center Online Service, provided
by the NASA/Goddard Space Flight Center.  KML gratefully acknowledges
support through NAG5-3307 and NAG5-7261 ({\it ASCA}) and NAG5-7971
(LTSA).

\appendix
\section{Appendix -- Examination of Excess Variance Assumptions}

As discussed in Section 3.2, three assumptions are implicitly made when
the excess variance is interpreted as the variability time scale. The
validity of two of the assumptions (that the length and sampling of
the light curves and that the slope of the power spectra are the same
for all objects) are examined in this section.  The other assumption,
that the normalization and slope of the power spectrum is not changing
within an observation,  is examined in
Section 3.3.1.

The excess variance can be related to the power spectrum of the
variability by the following equation:

\begin{displaymath}
\left<(h-\bar h)^2\right > \approx \int_{2\pi/T}^{2\pi/T_{bin}} P(f)
W(f) df
\end{displaymath}
where $f$ is the frequency, $P(f)$ is the variability power spectrum,
$W(f)$ is the window function, $T$ is the length of the observation,
and $T_{bin}$ is the bin size (e.g.\ Press \& Rybicki\markcite{142}
1997).

The steep frequency dependence of the AGN power spectrum means that
there should be a strong dependence on the span of the observation
$T_{obs}$.  For $\alpha=1.5$ and $2$ the dependence goes approximately
as $\sqrt(T_{obs})$ and $T_{obs}$, respectively.  Therefore,
comparison of the excess variances is only valid to the extent of that
the observations are the same length.  For the NLS1 sample, the
shortest and longest time spans are 21~ks and 200~ks for
IRAS~13349+2438 and PHL~1092 respectively.  Thus for $\alpha=1.5$ and
$2.0$ power laws, the potential difference in excess variance are
factors of 3 and 9.5 respectively, for this extreme case.  This
estimate changes somewhat when the redshift of the source is taken
into account, effectively reducing the spans by $1/(1+z)$.  Then the
shortest and longest time spans are 19~ks and 158~ks.  However, the
distribution of the square root of the spans (appropriate if
$\alpha=1.5$) is sharply peaked with average and standard deviation of
$9.5 \pm 2.1 \rm\, ks^{1\over 2}$, (corresponding to 90~ks) and 8 of 25
spans lie outside of this range.  With redshifts taken into account, the
result changes only slightly, to $9.1 \pm 1.9 \rm \, ks^{1\over 2}$,
with 7 lying outside of this range.

The other concern is that the window function should be the same for
all observations.  This is approximately true, but there are some
differences, originating for example in whether or not the SAA passage
coincides with the earth occultation.  The ratio of the
exposure to the span varies from 0.31 to 0.59 with average and
standard deviation of $0.41\pm 0.066$ and 8 outside this range.

The final implicit assumption required for a direct comparison of the
excess variances is that the shape of the power spectrum should be the
same in all objects.  This is difficult to evaluate directly because
of the gaps in the light curves.  Analysis methods exist which can
account for the gaps directly directly (e.g.\ Merrifield \&
McHardy\markcite{108} 1994; Papadakis \& McHardy\markcite{132} 1995;
Yaqoob et al.\markcite{180} 1997b); however, I use a Monte Carlo
approach adapted from Done et al.\markcite{33} 1992.  Simulations were
made for the 6 observations of 5 objects in which the skew parameter
computations could distinguish between variability power indices
(Section 3.3.2).  Simulated light curves are made with known $\alpha$
and no low frequency cutoff and the same sampling and measurement
noise properties of the real data.  Then the logarithmically rebinned
Lomb periodogram (e.g.\ Press et al.\markcite{143} 1992) of the
original data is compared with multiple examples of the simulated
data.  A potential problem with this approach is that the periodogram
is not a particularly good estimator of the power spectral index for
red noise power spectra (e.g.\ Pilgram \& Kaplan\markcite{136} 1998).
A better estimate might be obtained if one works in the time domain
and computes the structure or autocorrelation function as that is
likely to have better convergence properties; therefore the unbinned
discrete autocorrelation function (DCF; Edelson \& Krolik\markcite{37}
1988) was also used.

Done et al.\markcite{33} (1992) computed the Lomb periodogram for 20
simulated light curves with a range of $\alpha$, found the the mean
and standard deviation at each $\alpha$ and compared this with the
result from the real light curve.  A potential problem with this
approach is that the inherent nonstationarity of light curves with
$1/f$ power spectra may cause problems with convergence.  To test
whether or not the periodogram and DCF converge, ten sets of 100
simulations were generated for each test $\alpha$.  For each set of
100 simulations, the average and variance of the Lomb periodogram and
DCF were calculated.  These were then compared with the real data by
computing $\chi^2$.  The average and standard deviation of the 10
$\chi^2$ are plotted in Figure 9.  No low frequency cutoff was
assumed for the power spectrum of the simulated light curves.  The
effect of a low frequency cutoff would be to move the $\chi^2$ minimum
to higher values of $\alpha$.  There is clearly a convergence for each
index value, and it is clear that the convergence is faster for
smaller values of $\alpha$ than for larger ones.  This makes sense
since for large values of $\alpha$ there is a higher emphasis on low
frequencies which, because of the finite length of the light curve,
are not well constrained in a single light curve. The agreement
between the periodogram and the DCF methods appears to be good.

These plots show that there is a $\chi^2$ minimum between $\alpha=1$
and 2 for all objects, and therefore it appears that the assumption
that all light curves have the same power spectra is a good one.  Note
however that an implicit assumption is that a power law provides a
good fit to the periodogram and the DCF.  It is difficult to test this
assumption explicitly using $\chi^2$ goodness of fit criteria, because
adjacent points in the light curve are coupled and therefore, we do
not know how many degrees of freedom there are.  However, assuming
that this model is correct and does give a good fit, we can estimate
the best fit $\alpha$ and obtain rough estimates of the uncertainties.
The curves in Fig.\ 9 were fitted with a cubic model to obtain the
minimum which is then taken to be the best fit $\alpha$ (Table 3).
The error bars on the best fit were evaluated by determining what the
difference in $\alpha$ would be to find a $\chi^2$ value such that the
increase in $\chi^2$ is more than the sum of the $\chi^2$ uncertainty
on that value and the average of the five uncertainties surrounding
the best fit value.  Values beyond the range defined by the error bars
are unlikely to be obtained in an average of 100 simulations.

\section{Appendix -- Detection of Nongaussianity and Nonlinearity}

Detection of non-Gaussianity in lightcurves with $1/f$-type power
spectra is possible using parameters based on the skew and kurtosis,
or the 3rd and 4th moments of the flux distribution.  
 Lomnicki\markcite{98} (1961) shows that for
large N, two statistics $G_1(var G_1)^{-{1\over 2}}$ and $G_2 (var
G_2)^{-{1\over 2}}$ have approximately normal distributions with zero
mean and variance equal to 1, where $G_1=M_3 M_2^{-{3\over2}}$,
$G_2=M_4 M_2^{-2} -3$, $var G_1=6
\sum_{q=-\infty}^{q=+\infty} \rho_q^3$, and $var G_2=24
\sum_{q=-\infty}^{q=+\infty} \rho_q^4$, where $\rho_k$ are the
autocorrelation coefficients derived from the data, and the $M$ are
the central moments $M_r=N^{-1} \sum_{t=1}^{N} (x_t-m_1)^r$, here for
$r=2,3,4$ and $m_1$ is the mean.  It can be easily seen that $G_1$ and
$G_2$ are related to the skew and kurtosis; being independent
statistics, in principle either or both can be used.

The problem with using this formalism directly is that the summations
required to determine $var G_1$ and $var G_2$ cannot be evaluated if
the variability power spectrum is a power law with slope between 1 and
3, because, as discussed by Press \& Rybicki\markcite{142} (1997), the
variance is infinite and the mean is undefined.  Press \& Rybicki
advocate use of three-point statistics which possess a certain
symmetry and which are shown to behave well in  these circumstances.
Unfortunately, three-point statistics cannot be applied to these noisy
data.  Therefore, the skew and kurtosis approach is used, and the
variance is evaluated using Monte-Carlo simulations.  The values of
the skew and kurtosis are listed in Table 2.  The errors given are
based on the assumption of uncorrelated $e_t$ and therefore represent
lower limits.  I simulated 1000 light curves for each object
constructed to have a power law power spectrum with $\alpha=-1.5$;
these were then resampled to have the same length and window function
as the real, non-background subtracted light curves.  They were
renormalized to have the same mean and {\it true} variance as the real
data, and finally appropriate Poisson noise was added.  Finally, skew
and kurtosis of the simulated light curves were calculated. The
simulated light curves were constructed using the usual inverse
Fourier transform method; i.e.\ Fourier amplitudes with a power law
shape were computed, then the components were summed up with random
phases (e.g.\ Done et al.\markcite{33} 1992).  Note that this is
possibly the only way to do this in order to test against the
hypothesis of a correlated Gaussian power-law power spectrum.
Alternative methods such as the sum-of-shots method advocated by
Papadakis \& Lawrence\markcite{131} (1995) will not produce a Gaussian
simulated light curve, since the input components are non-Gaussian.

For each object, the distribution of the skews from the 1000 simulated
light curves is a good approximation of a normal distribution;
therefore, the mean and standard deviation of the simulated skews were
calculated.  As expected, those parameters depend on the particular
object: the peak is offset toward positive values when the fraction of
Poisson noise is high, as expected, since a Poisson distribution is
skewed, and the width of the distribution is larger for longer
observations.

A ``skew parameter'' was constructed from the simulations.  In order
to remove the skewness due to Poisson noise, the mean of the skews
from the simulated data was subtracted from the measured skew for each
object.  The result was divided by the standard deviation of the
distribution.  Thus the skew parameter is essentially the number of
$\sigma$ significance from being unskewed.  This value of the skew
parameter is listed in column 6 of Table 2.

As discussed above, the skew parameter $G_1$ is expected to depend on
the correlations present in the data.  If there are strong, long time
scale correlations, the observed skew will be larger.  To evaluate
this dependence, two more sets of simulations with power spectral
indices $\alpha=-1$ and $-2$ were performed for each object, and the
distribution of skews from these simulations were used to estimate
uncertainty on the skew parameter.  For many of the shorter and/or
less variable light curves, there was little difference in between
skew distributions for all three values of $\alpha$.  For these
objects the best guess on the uncertainty in the skew parameter must
be the value for uncorrelated data: $\sqrt(6/N)$ where $N$ is the
number of points.  For these observations, the uncertainty is near 1,
as would be expected if the data are uncorrelated.  The following
observations showed strong enough variability and/or good enough
statistics that the skew could distinguish between the three values
$\alpha$: NGC~4051 (2), IRAS~13224$-$3809 and Akn~564.  For these
objects, the derived skew is significantly larger if the underlying
power spectrum has $\alpha=-1$ compared with $-2$.  The following
observations could distinguish between $\alpha=-1.5$ and $-2$, but not
between $-1$ and $-1.5$: Ton~S180, PHL~1092, 1H~0707$-$495, NGC~4051
(1).

The kurtosis of the light curves was computed and it is listed in the
7th column of Table 2.  It is much less straightforward to derive a
statistic from the real data and the results of the simulations
because the kurtosis appears to be directly affected by the long range
correlations in the data, and the distribution of the kurtosis values
is distinctly not normal. However, it is clear that in the cases that
skew parameter is large, the kurtosis is large and positive,
supporting the idea that those light curves are non-Gaussian.

\newpage

\begin{deluxetable}{lllllllll}
\scriptsize
\tablewidth{0pc}
\tablenum{1}
\tablecaption{Observation Log}
\tablehead{
\colhead{Target} & \colhead{Sequence} 
& \colhead{Observation} &
\colhead{Observing} & \colhead{Exposure} & \colhead{Time}
& \colhead{Count\tablenotemark{1}}  & 
 \colhead{Reference} \\
& \colhead{Number} & \colhead{Start Date} & \colhead{Mode} & & \colhead{Span}
& \colhead{Rate} & \colhead{Rate}\\
& & &  & \colhead{(ks)} & \colhead{(ks)}
& \colhead{(cnts/s)} & \\}
\startdata

Mrk 335 & 71010000 & 93.343 & BRIGHT(2) & 19.4 & 46.9 & $1.287$  & N97a, N97b, R97 \nl
I Zw 1 & 73042000 & 95.196 & FAINT(1) & 28.3 & 92.4 & $0.403$ &  \nl
Ton S180 & 74081000 & 96.192 & FAINT(1) & 47.9 & 94.3 & $0.982$ \nl
PHL 1092 & 75042000 & 97.197 & FAINT(1) & 72.2 & 199.6 & $0.039$ \nl
RX J0439$-$45 & 75050000 &  97.289 & FAINT(1) & 45.7 & 139.8 & $0.163$ \nl
NAB 0205+024 & 74071000 & 96.018 & BRIGHT(2) & 51.8 & 118.6 &
$0.350$ & F98 \nl
PKS 0558$-$504 & 74096000 & 96.249 & FAINT(1) & 35.7 & 94.7 & $1.74$ \nl
1H 0707$-$495 & 73043000 & 95.074 & FAINT(1) & 38.7 & 97.8 & $0.194$ & L97b \nl
Mrk 142 & 76034000 & 98.124 & FAINT(1) & 19.3 & 48.9 & $0.352$ & \nl
RE 1034+39 & 72020000 & 94.323 & FAINT(1) & 30.0 & 88.5 & $0.187$ & PDO95 \nl
NGC 4051(1) & 70001000 & 93.115 & BRIGHT(4) & 31.6 & 82.1 & $2.493$ & M94\nl
NGC 4051(2) & 72001000 & 94.158 & FAINT(1) & 74.8 & 158.8 & $2.271$ & G96\nl
PG 1211+143 & 70025000 & 93.154 & BRIGHT(4) & 29.4 & 90.3 & $0.475$ & Y94 \nl
Mrk 766 & 71046000 & 93.352 & FAINT(1) & 34.4 & 77.4 & $1.608$ & L96,N97a,N97b \nl
PG 1244+026 & 74070000 & 96.183 & BRIGHT(2) & 38.3 & 108.3 & $0.493$ &  F98 \nl
IRAS 13224$-$3809 & 72011000 & 94.211 & FAINT(1) & 91.2 & 153.7 & $0.091$ & L97b \nl
IRAS 13349+2438(1) & 73056000 & 95.178 & FAINT(1) & 10.2 & 23.2 &
$0.668$ & B97; BME97 \nl
IRAS 13349+2438(2) & 73056010 & 95.181 & FAINT(1) & 7.6 & 20.9 &
$0.719$ & B97, BME97 \nl
PG 1404+226 & 72021000 & 94.194 & FAINT(1) & 35.1 & 96.2 & $0.080$ & L97b \nl
Mrk 478 & 73067000 & 95.183 & FAINT(1) & 33.3 & 84.5 & $0.297$ &  \nl
IRAS 17020+4544 & 73047000 & 95.241 & FAINT(1) & 36.5 & 84.1 &
$0.896$ & L97a \nl
Mrk 507 & 74033000 & 95.350 & FAINT(2) & 33.7 & 77.2 & $0.037$ & IBF98\nl
KAZ 163 & 74033000 & 95.350 & FAINT(2) & 33.7 & 77.2 & $0.059$ &   \nl
IRAS 20181$-$2244 & 73075000 & 95.284 & BRIGHT(1) & 56.0 & 113.2 &
$0.120$ & HM98 \nl
Akn 564 & 74052000 & 96.358 & FAINT(1) & 49.7 & 103.6 & $3.597$ \nl
\enddata
\tablenotetext{1}{Count rate is net for the sum of SIS0 and SIS1
detectors between 0.5 and 10.0 keV.}
\tablerefs{N97a: Nandra et al.\markcite{117} 1997a; N97b: 
Nandra et al.\markcite{118} 1997b; R97:
Reynolds\markcite{147} 1997; F98: Fiore et al.\markcite{44} 1998b;
 L97b: Leighly et al.\markcite{95} 1997b;
PDO95: Pounds, Done \& Osborne\markcite{140} 1995; M94:
 Mihara et al.\markcite{109} 1994; G96:
Guainazzi et al.\markcite{67} 1996; Yaqoob et al.\markcite{181}
 1994; L96: Leighly et al.\markcite{94} 1996;
B97: Brinkmann et al.\markcite{19} 1996; BME97: 
Brandt, Mathur \& Elvis\markcite{16} 1997;
L97a: Leighly et al.\markcite{78} 1997a; IBF98: 
Iwasawa, Brandt \& Fabian\markcite{75} 1998;
HM98: Halpern \& Moran\markcite{71} 1998}
\end{deluxetable}

\clearpage

\singlespace
\begin{deluxetable}{lllllll}
\scriptsize
\tablewidth{0pc}
\tablenum{2}
\tablecaption{Variability Properties}
\tablehead{
\colhead{Target} & \colhead{Variable $^1$} & \colhead{Variable $^2$}
 & \colhead{Excess} & \colhead{Skew} &
\colhead{Skew Parameter} & \colhead{Kurtosis}  \\
& \colhead{Segments} & \colhead{Segments} & \colhead{Variance} \\
& & & \colhead{($\times 10^{-2}$)} \\}
\startdata

Mrk 335 & 1/8 & 0/97 & $0.53 \pm 0.14$ & $0.19 \pm 0.22$ & $0.75 \pm
0.90$ & $-0.19 \pm 0.43$ \nl
I Zw 1 & 0/15 & 0/107 & $4.99 \pm 0.67$ & $0.08 \pm 0.18$ & $-0.34 \pm
0.65$ & $-0.36 \pm 0.37$  \nl
Ton S180 & 0/25 & 3/177 & $2.55 \pm 0.27$ & $0.73 \pm 0.14$ &
$2.63^{+1.25}_{-0.55}$ & $0.02 \pm 0.28$ \nl
PHL 1092 & 0/43 & 2/261 & $2.35 \pm 3.59$ & $1.02 \pm 0.11$ &
$5.43^{+1.32}_{-0.71}$ & $1.21 \pm 0.23$ \nl
RX J0439$-$45 & 0/22 & 0/106 & $4.04 \pm 0.79$ & $0.51 \pm 0.15$ & $2.21
\pm 0.84$ & $-0.07 \pm 0.31$ \nl
NAB 0205+024 & 0/24 & 0/198 & $2.02 \pm 0.29$ & $0.30 \pm 0.14$ &
$1.43 \pm 0.82$ & $-0.52 \pm 0.27$ \nl
PKS 0558$-$504 & 1/14 & 1/152 & $1.76 \pm 0.19$ & $0.08 \pm 0.16$ &
$0.24 \pm 0.51$ & $-0.32 \pm 0.33$ \nl
1H 0707$-$495 & 6/17 & 11/142 & $24.5 \pm 3.18$ & $1.02 \pm 0.16$ &
$3.19^{+1.33}_{-0.52}$ & $0.86 \pm 0.33$ \nl
Mrk 142 & 2/7 & 2/86 & $8.28 \pm 1.23$ & $0.31 \pm 0.22$ & $0.85 \pm
0.67$ & $-0.37 \pm 0.45$ \nl
RE 1034+39 & 1/17 & 6/121 & $2.42 \pm 0.69$ & $0.32 \pm 0.17$ & $1.17
\pm 0.94$ & $0.11 \pm 0.35$ \nl
NGC 4051(1) & 15/15 & 64/98 & $11.5 \pm 1.03$ & $0.47 \pm 0.18$ &
$1.21^{+0.58}_{-0.48}$ & $-0.63 \pm 0.37$ \nl
NGC 4051(2) & 28/34 & 161/283 & $19.17 \pm 1.48$ & $0.72 \pm 0.11$ &
$2.01^{1.54}_{-0.38}$ & $0.60 \pm 0.23$ \nl
PG 1211+143 & 0/15 & 0/72 & $1.26 \pm 0.47$ & $-0.05 \pm 0.19$ &
$-0.62 \pm 0.98$ & $-0.46 \pm 0.38$ \nl
Mrk 766 & 8/14 & 7/132 & $7.89 \pm 0.57$ & $-0.12\pm 0.17$ & $-0.34
\pm 0.48$ & $-1.06 \pm 0.34$ \nl
PG 1244+026 & 1/17 & 2/141 & $3.64 \pm 0.48$ & $0.02 \pm 0.16$ & $0.0
\pm 0.64$ & $-0.07 \pm 0.32$ \nl
IRAS 13224$-$3809 & 5/36 & 7/466 & $63.46 \pm 6.18$ & $1.65 \pm 0.10$ &
$5.66^{+4.19}_{-1.12}$ & $3.02 \pm 0.20$ \nl
IRAS 13349+2438(1) & 0/4 & 0/43 & $2.80 \pm 0.51$ & $0.59 \pm 0.30$ &
$1.79 \pm 0.96$ & $-0.89 \pm 0.61$ \nl
IRAS 13349+2438(2) & 0/5 & 0/19 & $0.61 \pm 0.26$ & $0.33 \pm 0.34$ &
$0.89 \pm 1.05$ & $-0.78 \pm 0.69$ \nl 
PG 1404+226 & 0/15 & 1/131 & $26.89 \pm 3.18$ & $0.62 \pm 0.17$ &
$2.11 \pm 0.62$ & $-0.37 \pm 0.34$ \nl
Mrk 478 & 0/18 & 0/115 & $1.52 \pm 0.41$ & $-0.05 \pm 0.17$ & $-0.73
\pm 0.93$ & $-0.02 \pm 0.33$ \nl
IRAS 17020+4544 & 1/13 & 4/155 & $1.42 \pm 0.20$ & $0.23 \pm 0.17$ &
$1.00 \pm 0.73$ & $-0.34 \pm 0.33$ \nl
Mrk 507 & 0/14 & 2/150 & $2.41 \pm 2.79$ & $0.61 \pm 0.17$ & $1.43 \pm
0.95$ & $0.58 \pm 0.33$ \nl
KAZ 163 & 0/14 & 0/151 & $2.36 \pm 1.56$ & $0.30 \pm 0.17$ & $0.21 \pm
0.99$  & $-0.34 \pm 0.33$ \nl
IRAS 20181$-$2244 & 1/29 & 0/218 & $3.30 \pm 0.71$ & $0.24 \pm 0.13$ &
$0.69 \pm 0.89$ & $-0.32 \pm 0.26$ \nl
Akn 564 & 17/26 & 68/205 & $3.97 \pm 0.43$ & $1.03 \pm 0.14$ &
$3.06^{+1.90}_{-0.67}$ & $1.36 \pm 0.27$ \nl
\enddata
\tablenotetext{1}{Number of variable continuous segments/total number
of continuous segments longer than 5 bins.}
\tablenotetext{2}{Number of variable segments 5 bins long/total number
of variable segments 5 bins long.  The
segments are not independent, so the number of detections should be
the total divided by $N-1=4$.}
\end{deluxetable}

\clearpage

\singlespace
\begin{deluxetable}{lllllll}
\normalsize
\tablewidth{0pc}
\tablenum{3}
\tablecaption{Power Spectral Indices}
\tablehead{
\colhead{Target} & \colhead{Index $^1$} & \colhead{Index $^1$} \\
& \colhead{(Lomb-Scargle)} & \colhead{(DCF)} \\}

\startdata
Ton S180 & $1.57^{+0.63}_{-0.47}$ & $1.55^{+0.35}_{-0.25}$ \\
PHL 1092 & $1.66^{+0.54}_{-0.46}$ & $1.84^{+0.28}_{-0.34}$ \\
NGC 4051 (1) & $1.41^{+0.19}_{-0.31}$ & $1.26^{+0.24}_{-0.26}$ \\
NGC 4051 (2) & $1.15 \pm 0.15$ & $1.16^{+0.24}_{-0.26}$ \\
IRAS 13224-3809 & $1.52^{+0.28}_{-0.22}$ & $1.53^{+0.17}_{-0.13}$ \\
Ark 564 & $1.28^{+0.12}_{-0.28}$ & $1.34^{+0.36}_{-0.24}$ \\

\enddata
\tablenotetext{1}{These values are estimated from the simulations
shown in Figure 9 (see text).}
\end{deluxetable}

\clearpage

\renewcommand{\thefigure}{\arabic{figure}}
\setcounter{figure}{0}

\begin{figure}[t]
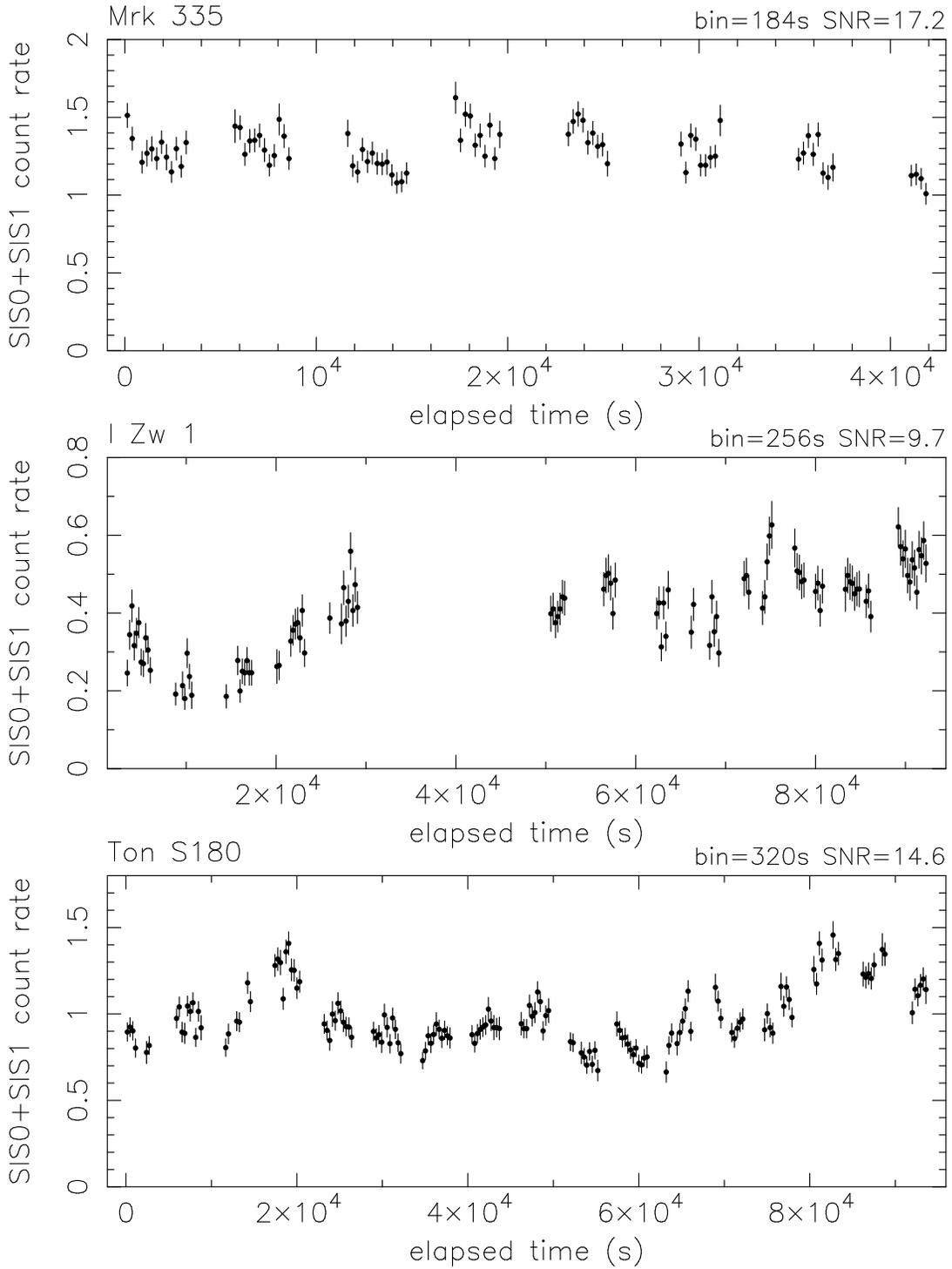

\vbox to2.5in{\rule{0pt}{2.5in}}
\includegraphics{fig1a.ps}
\vbox to2.5in{\rule{0pt}{2.5in}}
\includegraphics{fig1b.ps}
\vbox to2.5in{\rule{0pt}{2.5in}}
\includegraphics{fig1c.ps}
\caption{SIS0+SIS1 light curves from {\it ASCA} observations of
Narrow-line Seyfert 1 galaxies.  For the purpose of display, range of
bin sizes were used to account for the range of fluxes and time scales
of variability present.  The exposure parameter (fraction of
incomplete bins included) ranged from 0.5 to 1.  The bin size and
average signal to noise is listed.}
\end{figure}

\newpage

\renewcommand{\thefigure}{\arabic{figure}}
\setcounter{figure}{0}

\begin{figure}[t]
\vbox to2.5in{\rule{0pt}{2.5in}}
\includegraphics{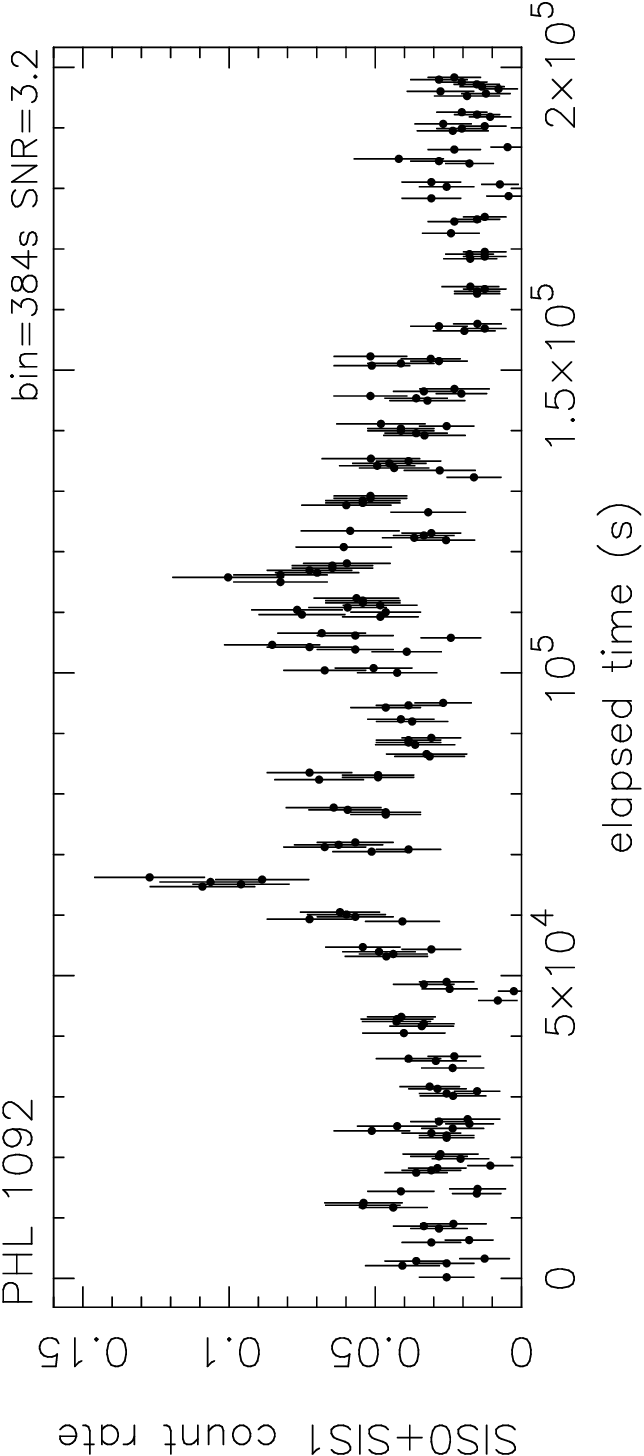}
\vbox to2.5in{\rule{0pt}{2.5in}}
\includegraphics{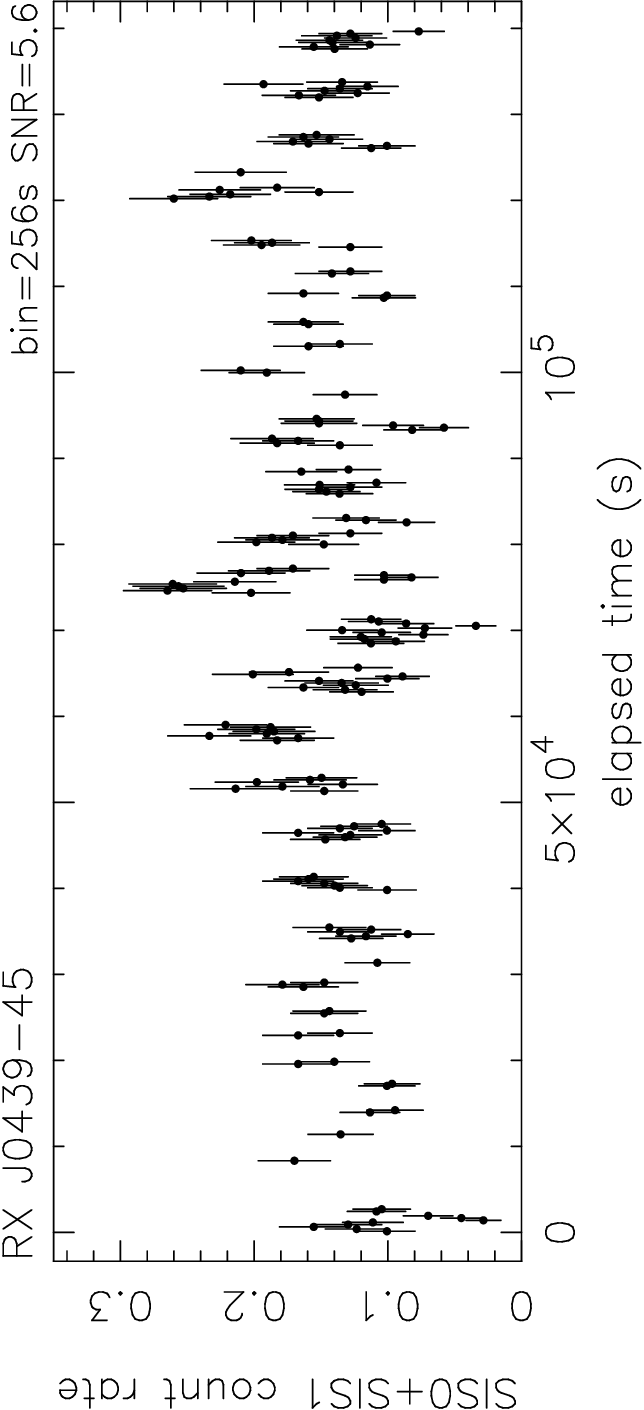}
\vbox to2.5in{\rule{0pt}{2.5in}}
\includegraphics{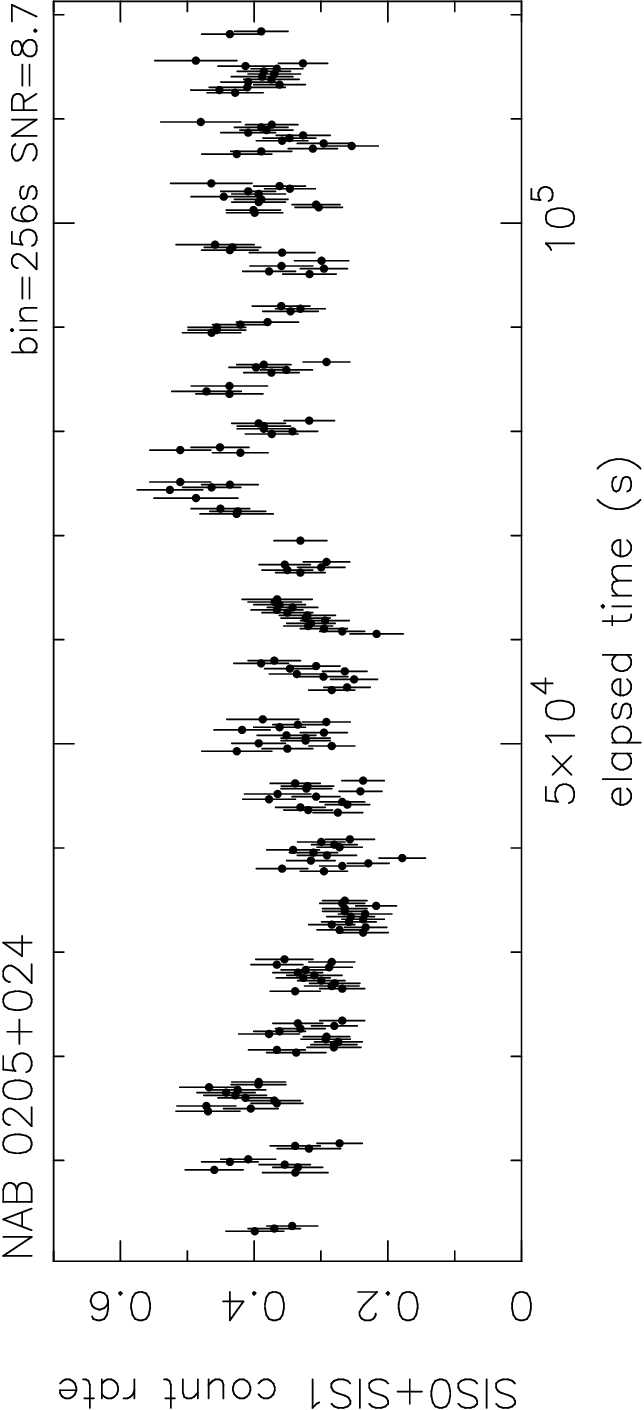}
\caption{continued.}
\end{figure}

\newpage

\renewcommand{\thefigure}{\arabic{figure}}
\setcounter{figure}{0}

\begin{figure}[t]
\vbox to2.5in{\rule{0pt}{2.5in}}
\includegraphics{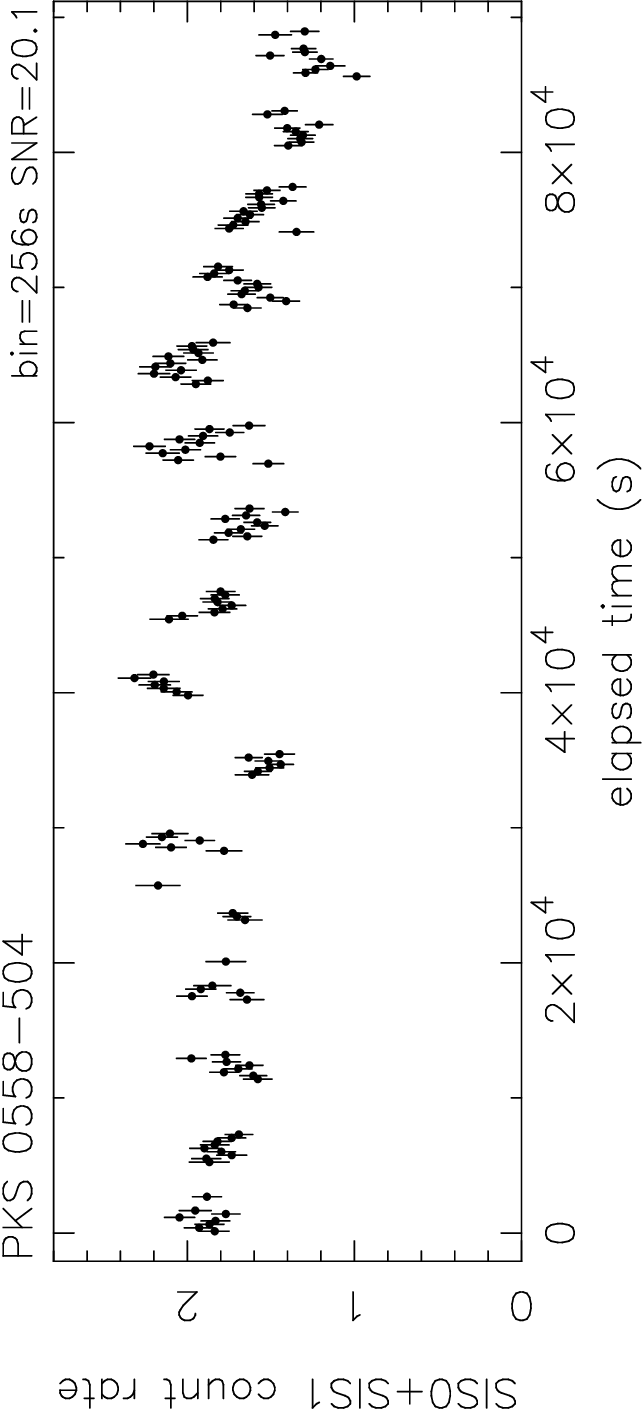}
\vbox to2.5in{\rule{0pt}{2.5in}}
\includegraphics{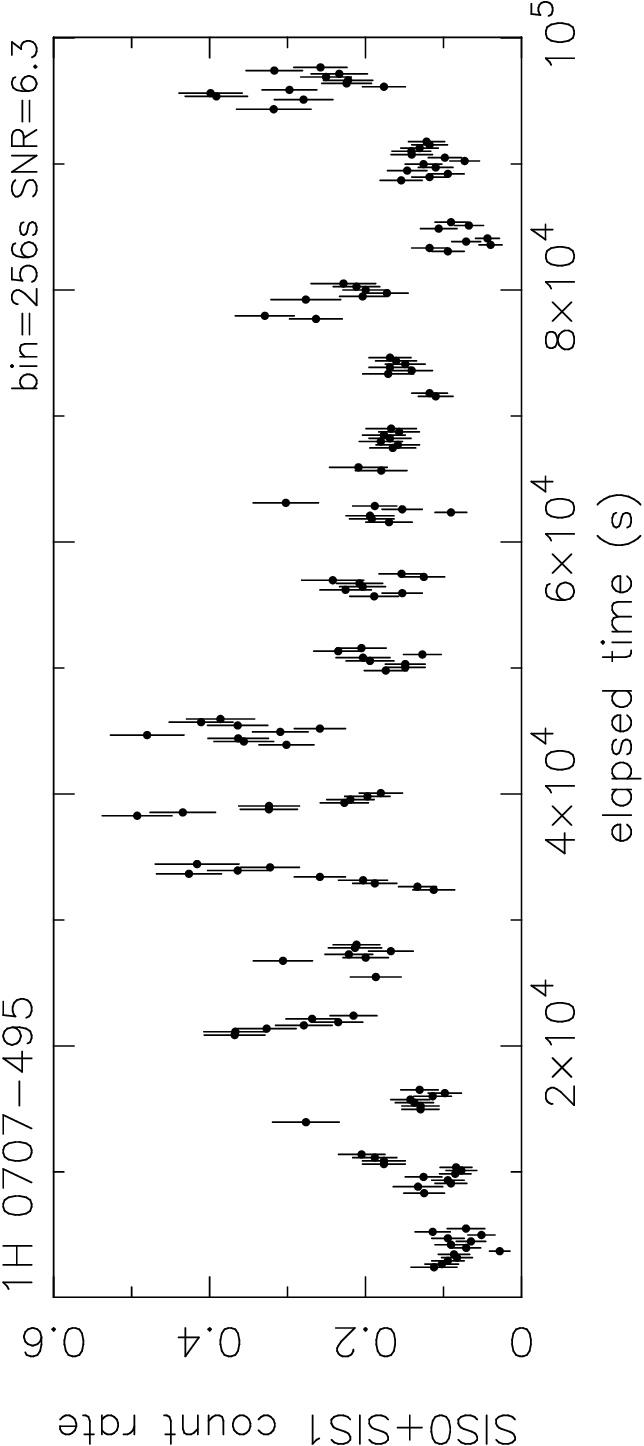}
\vbox to2.5in{\rule{0pt}{2.5in}}
\includegraphics{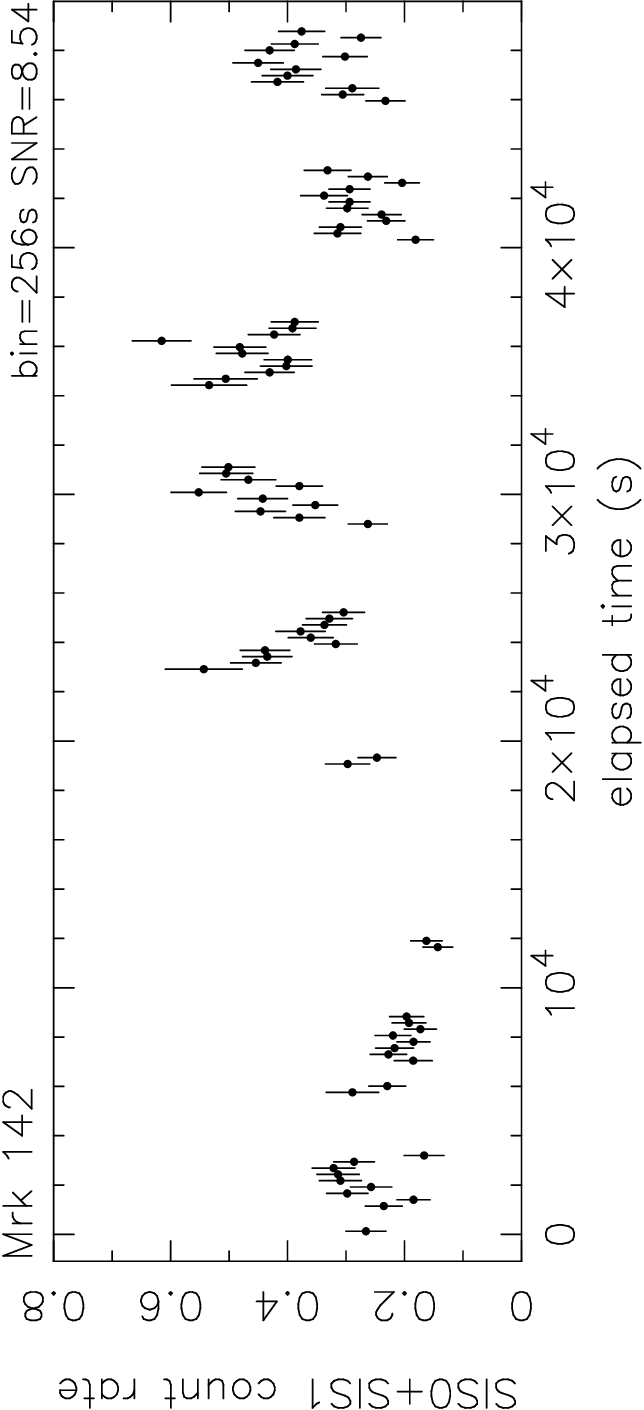}
\caption{continued.}
\end{figure}

\newpage

\renewcommand{\thefigure}{\arabic{figure}}
\setcounter{figure}{0}

\begin{figure}[t]
\vbox to2.5in{\rule{0pt}{2.5in}}
\includegraphics{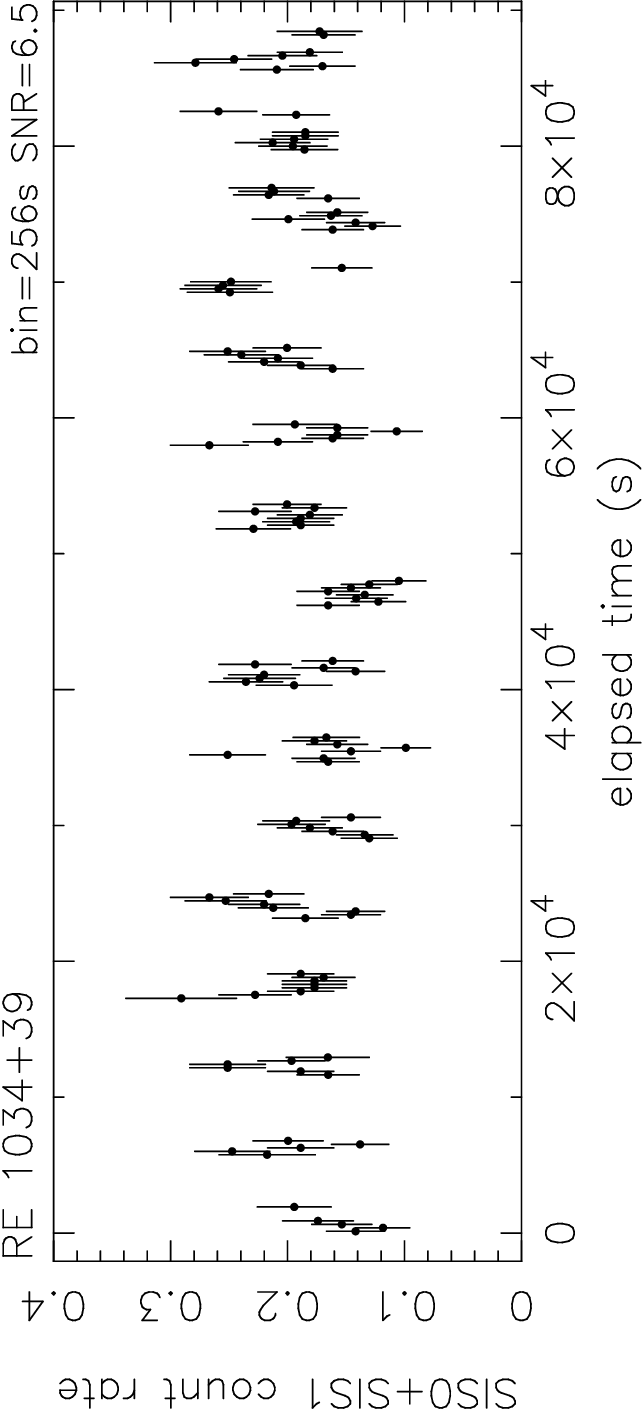}
\vbox to2.5in{\rule{0pt}{2.5in}}
\includegraphics{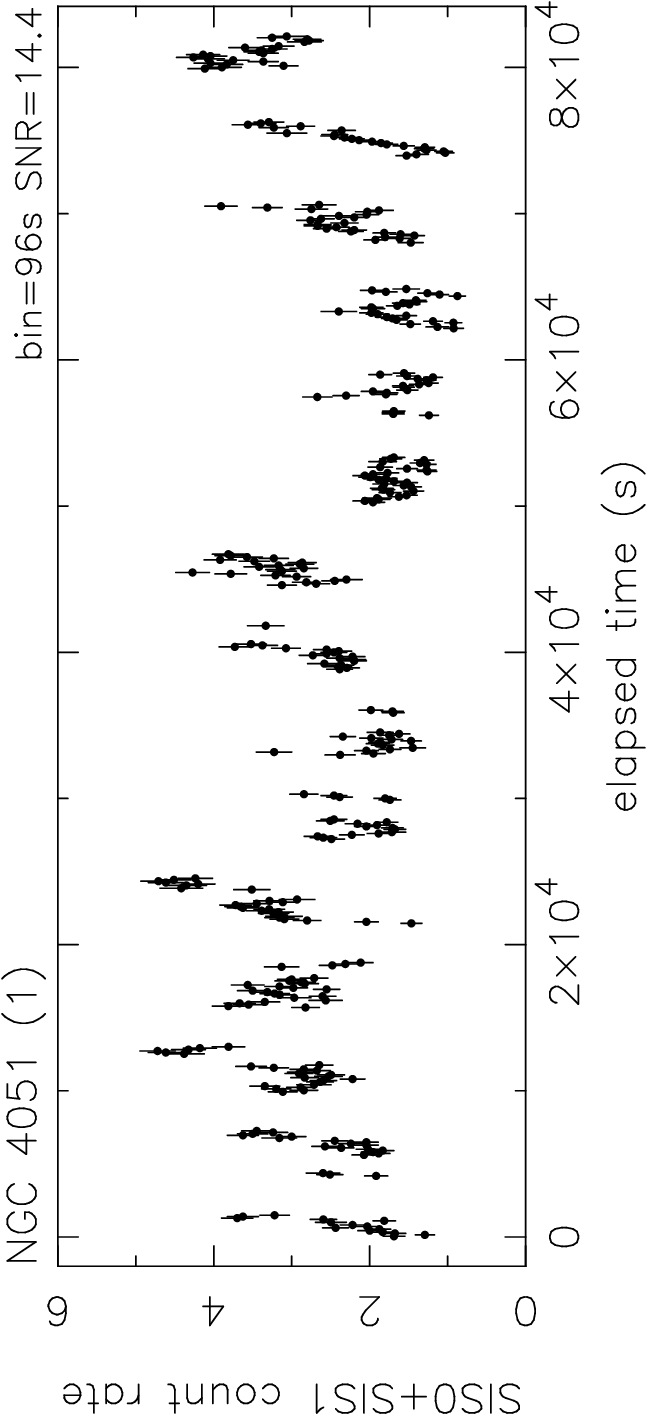}
\vbox to2.5in{\rule{0pt}{2.5in}}
\includegraphics{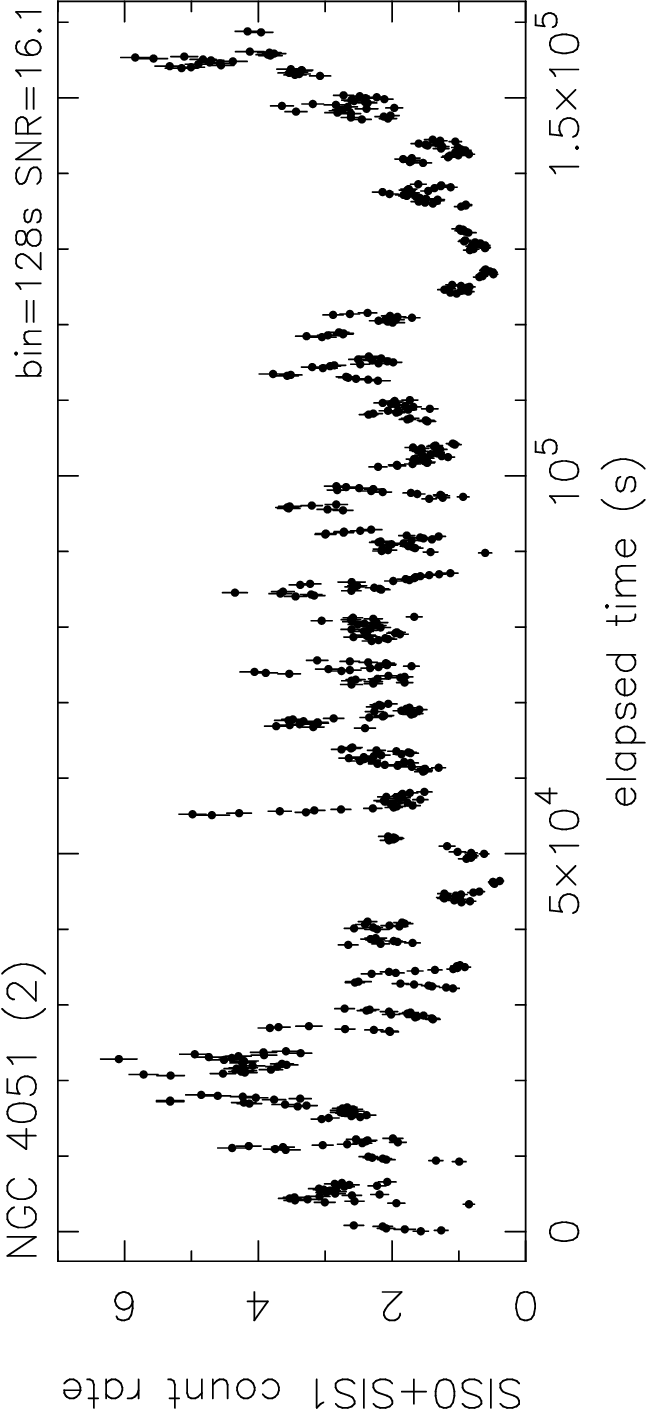}
\caption{continued.}
\end{figure}

\newpage

\renewcommand{\thefigure}{\arabic{figure}}
\setcounter{figure}{0}

\begin{figure}[t]
\vbox to2.5in{\rule{0pt}{2.5in}}
\includegraphics{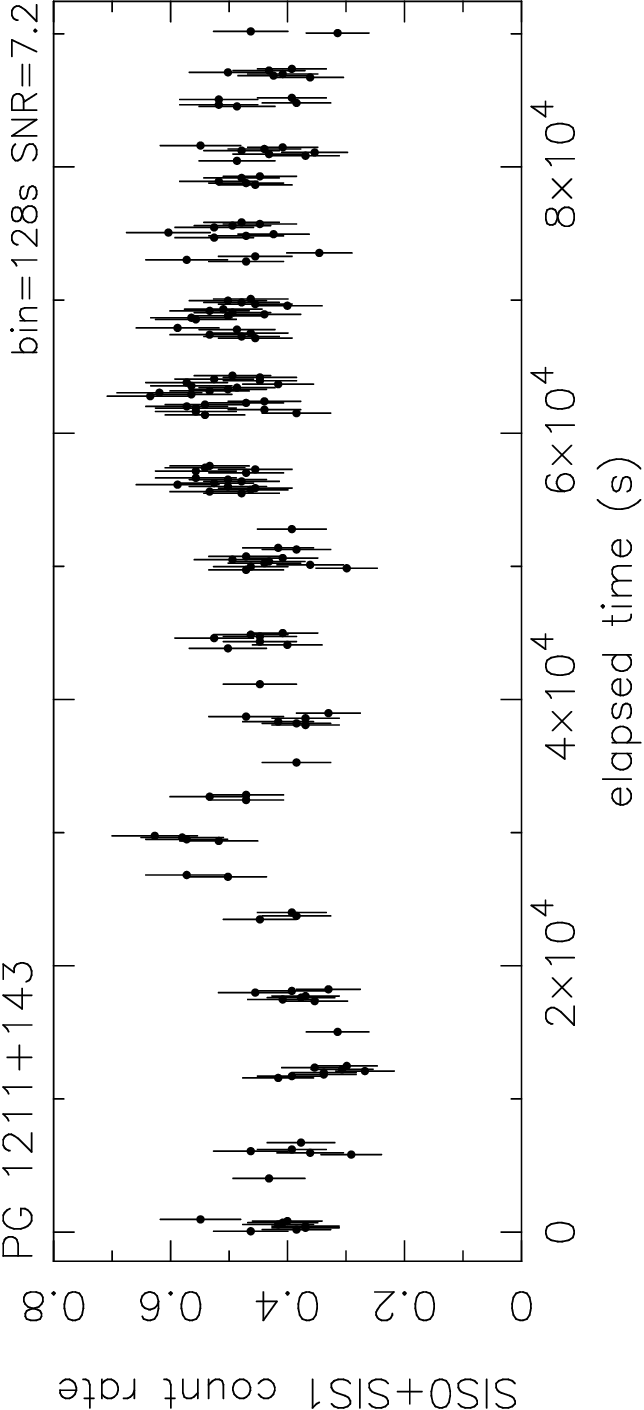}
\vbox to2.5in{\rule{0pt}{2.5in}}
\includegraphics{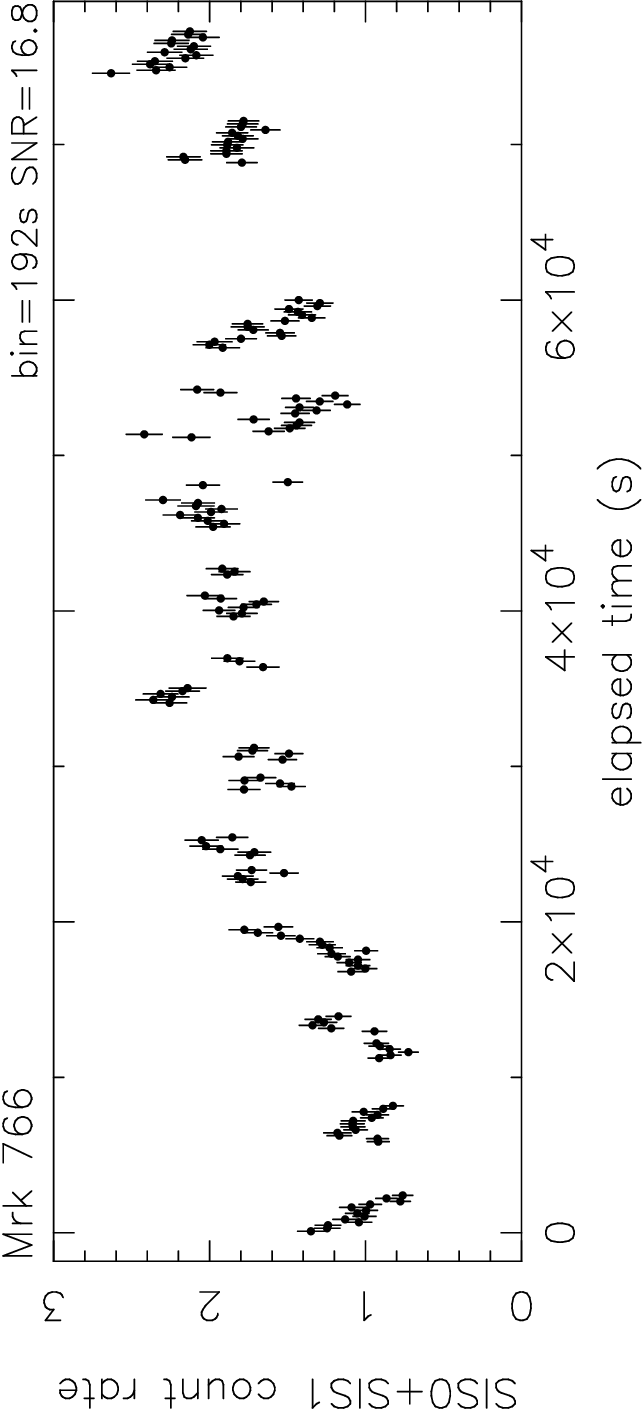}
\vbox to2.5in{\rule{0pt}{2.5in}}
\includegraphics{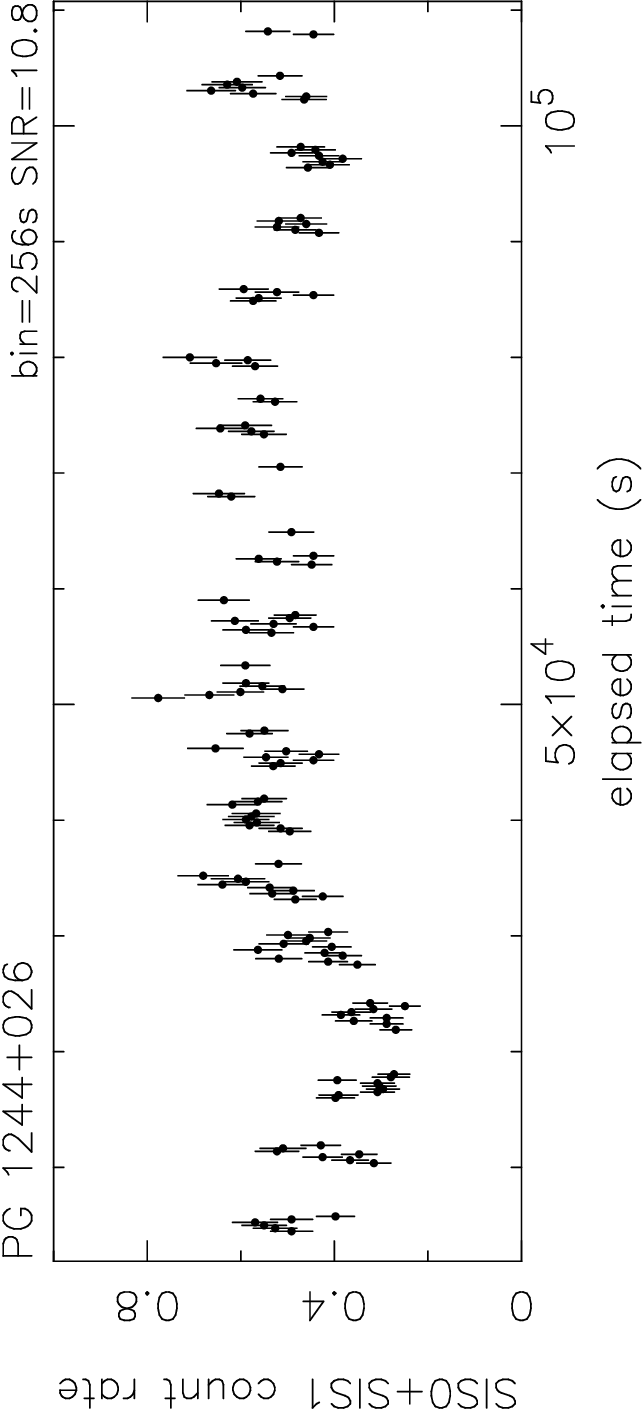}
\caption{continued.}
\end{figure}

\newpage

\renewcommand{\thefigure}{\arabic{figure}}
\setcounter{figure}{0}

\begin{figure}[t]
\vbox to2.5in{\rule{0pt}{2.5in}}
\includegraphics{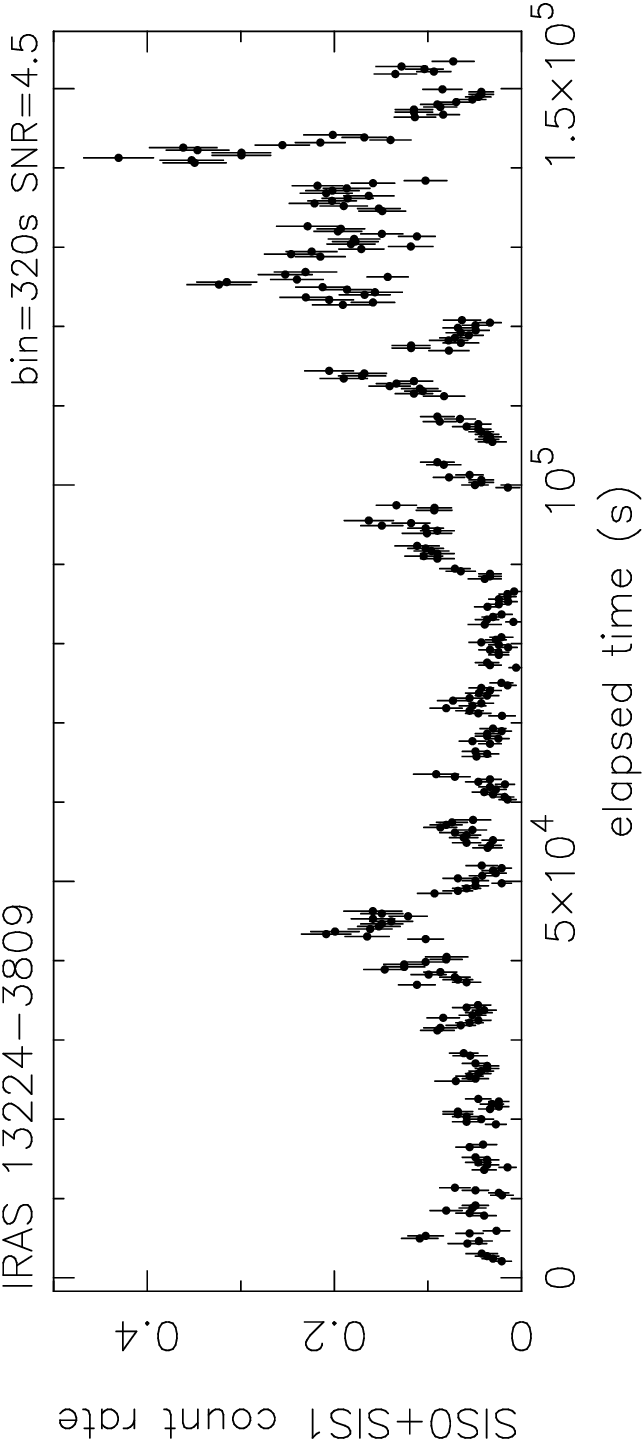}
\vbox to2.5in{\rule{0pt}{2.5in}}
\includegraphics{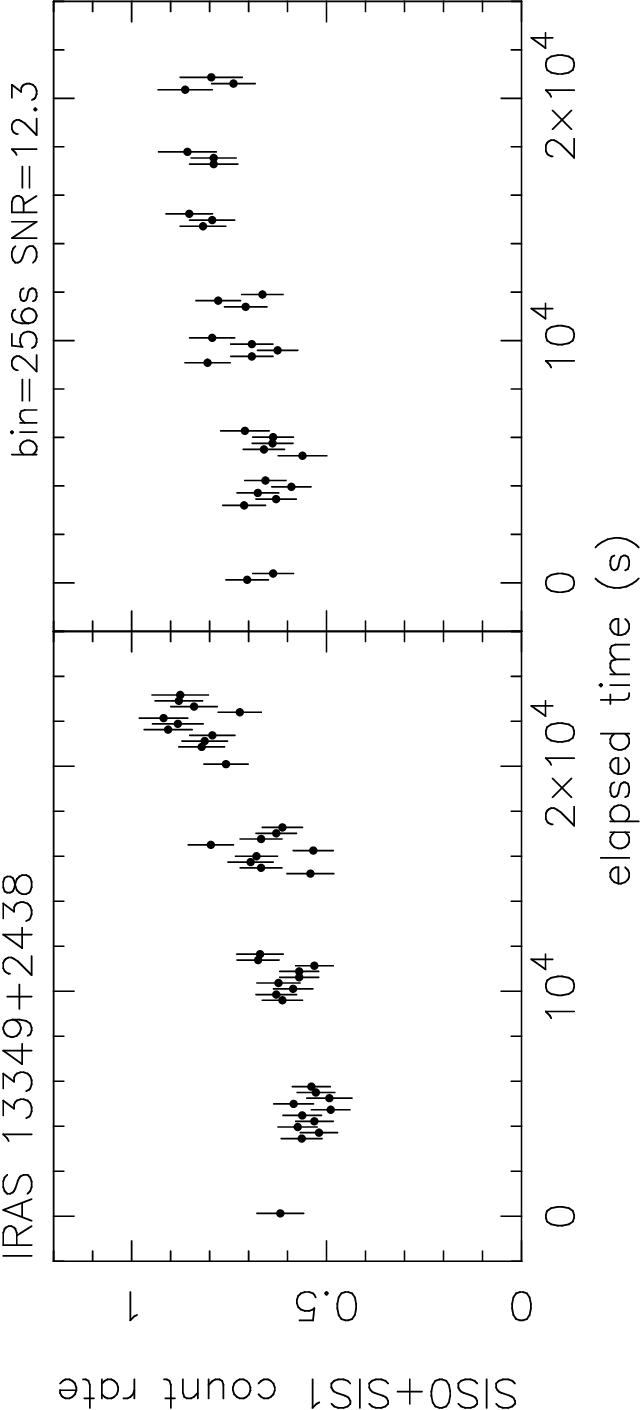}
\vbox to2.5in{\rule{0pt}{2.5in}}
\includegraphics{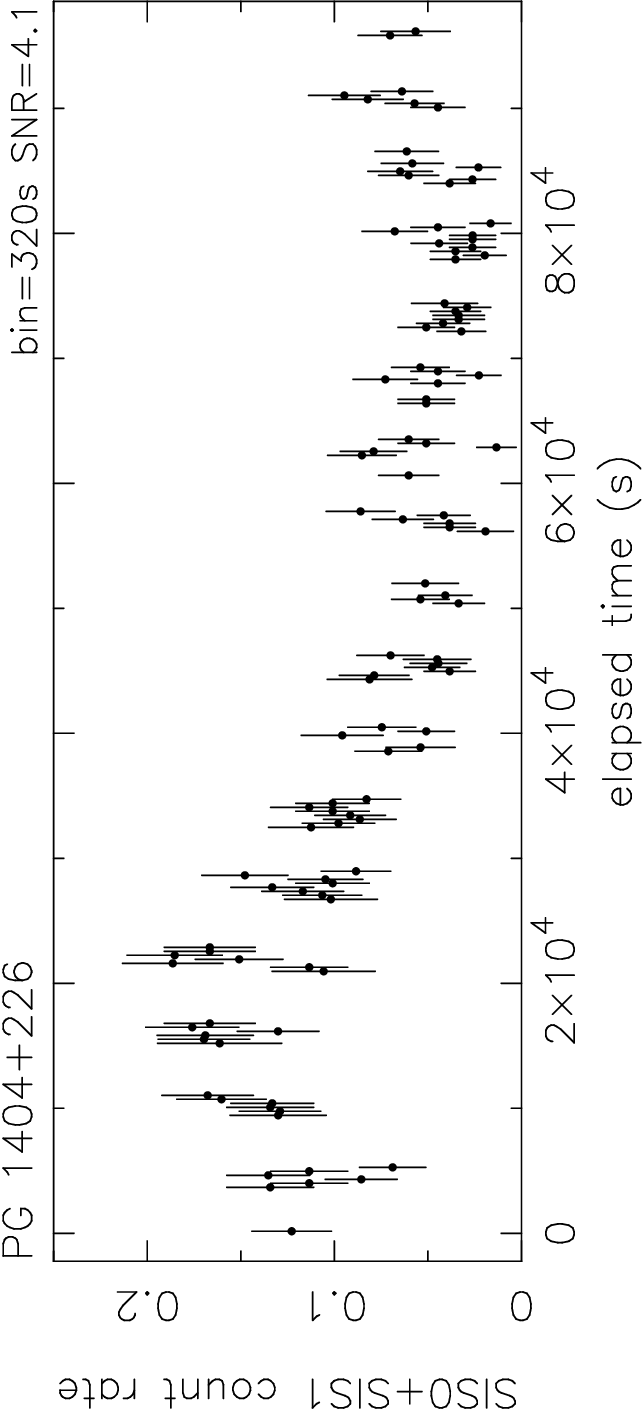}
\caption{continued.}
\end{figure}

\newpage

\renewcommand{\thefigure}{\arabic{figure}}
\setcounter{figure}{0}

\begin{figure}[t]
\vbox to2.5in{\rule{0pt}{2.5in}}
\includegraphics{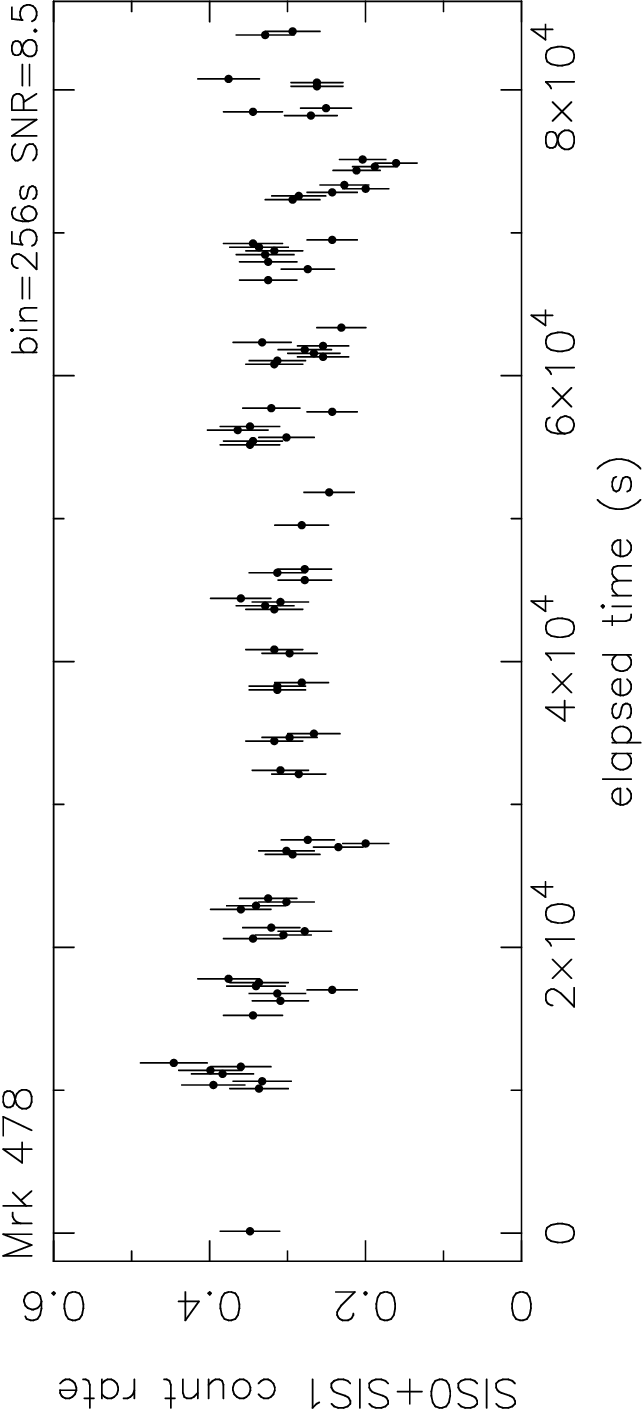}
\vbox to2.5in{\rule{0pt}{2.5in}}
\includegraphics{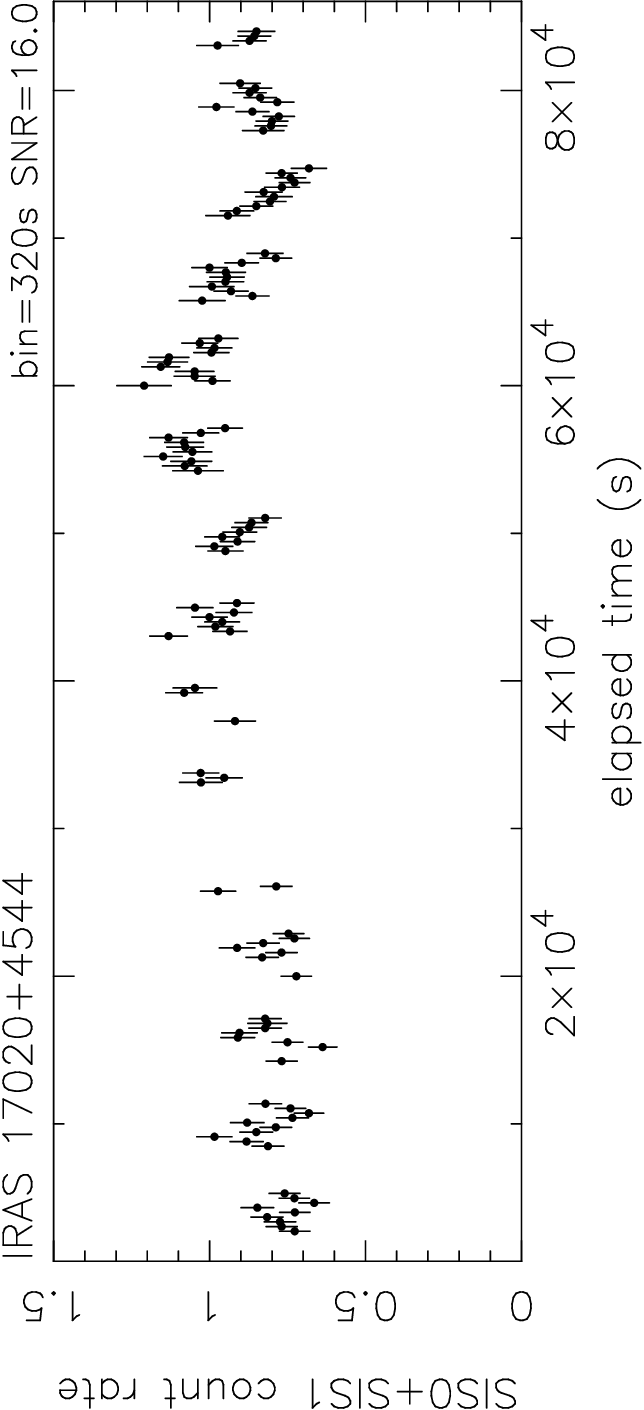}
\vbox to2.5in{\rule{0pt}{2.5in}}
\includegraphics{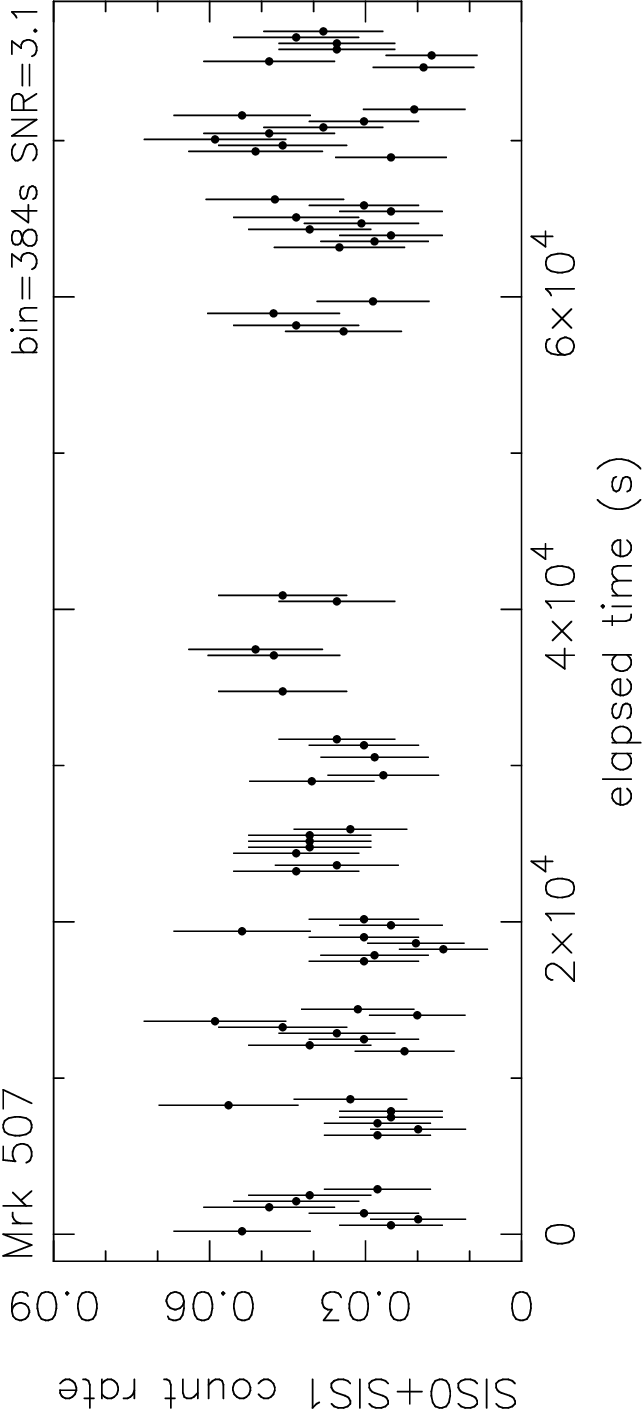}
\caption{continued.}
\end{figure}

\newpage

\renewcommand{\thefigure}{\arabic{figure}}
\setcounter{figure}{0}

\begin{figure}[t]
\vbox to2.5in{\rule{0pt}{2.5in}}
\includegraphics{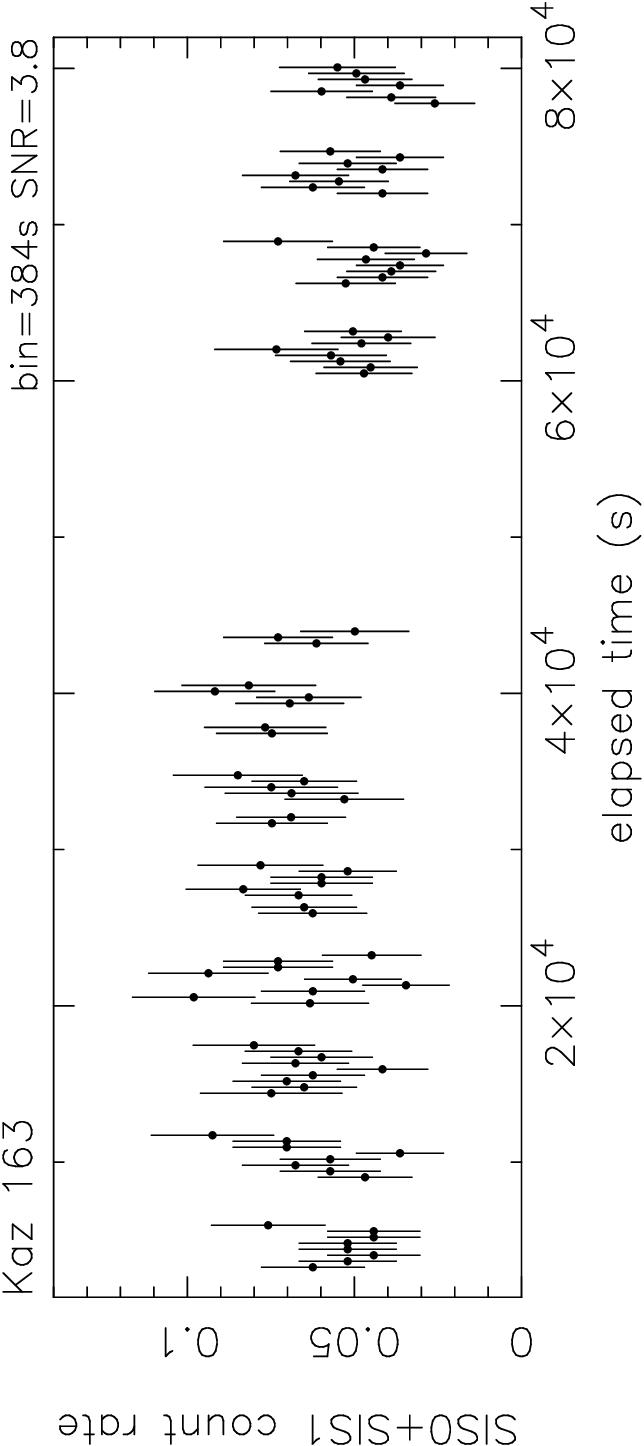}
\vbox to2.5in{\rule{0pt}{2.5in}}
\includegraphics{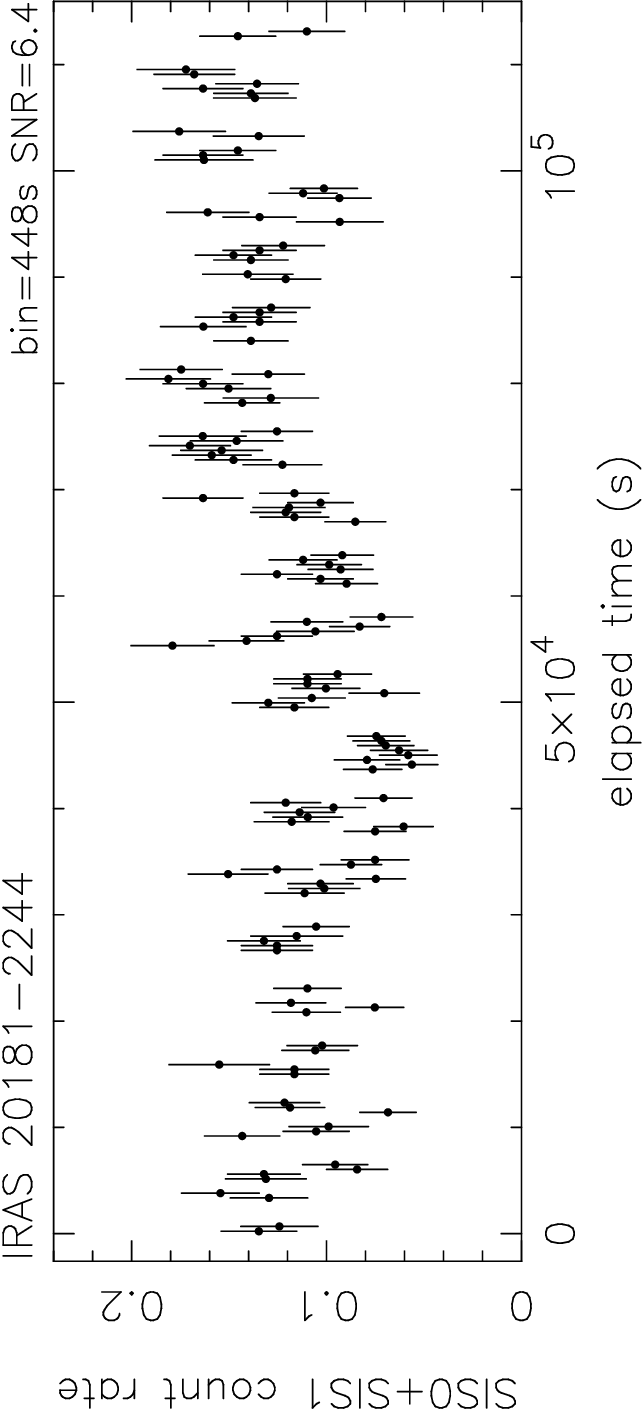}
\vbox to2.5in{\rule{0pt}{2.5in}}
\includegraphics{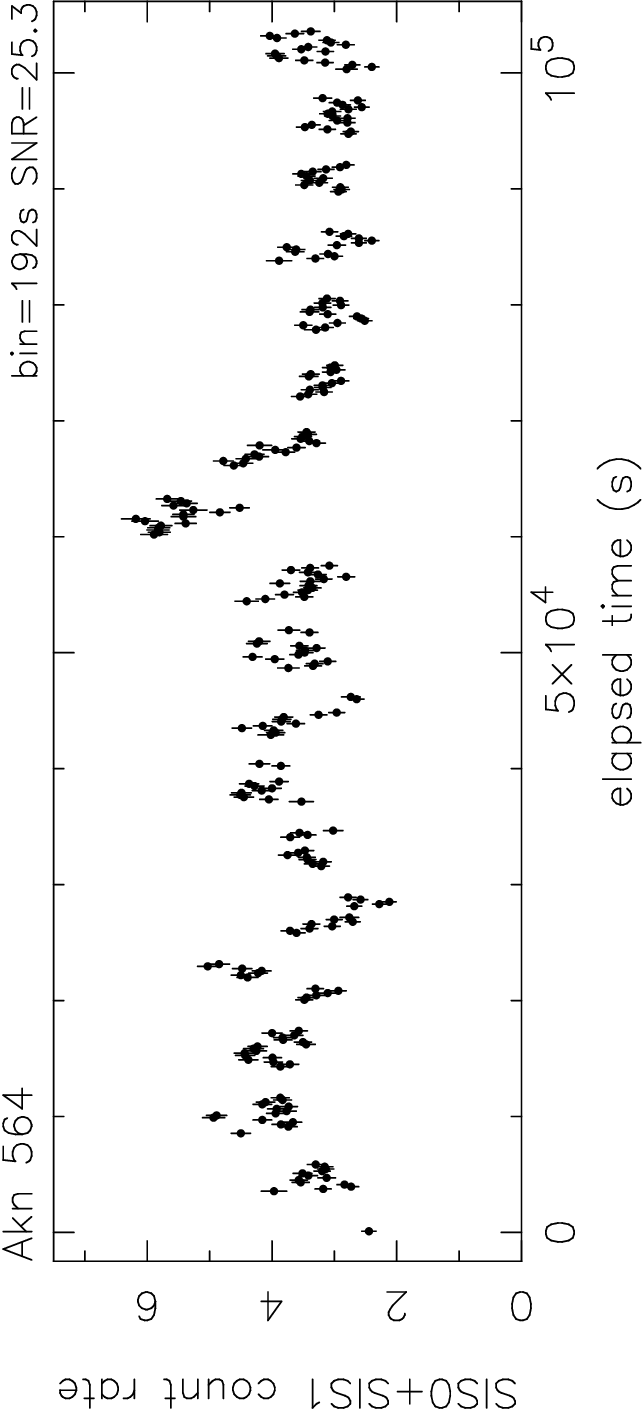}
\caption{continued.}
\end{figure}

\newpage

\renewcommand{\thefigure}{\arabic{figure}}
\setcounter{figure}{1}

\begin{figure}[t]
\vbox to3.5in{\rule{0pt}{3.5in}}
\includegraphics{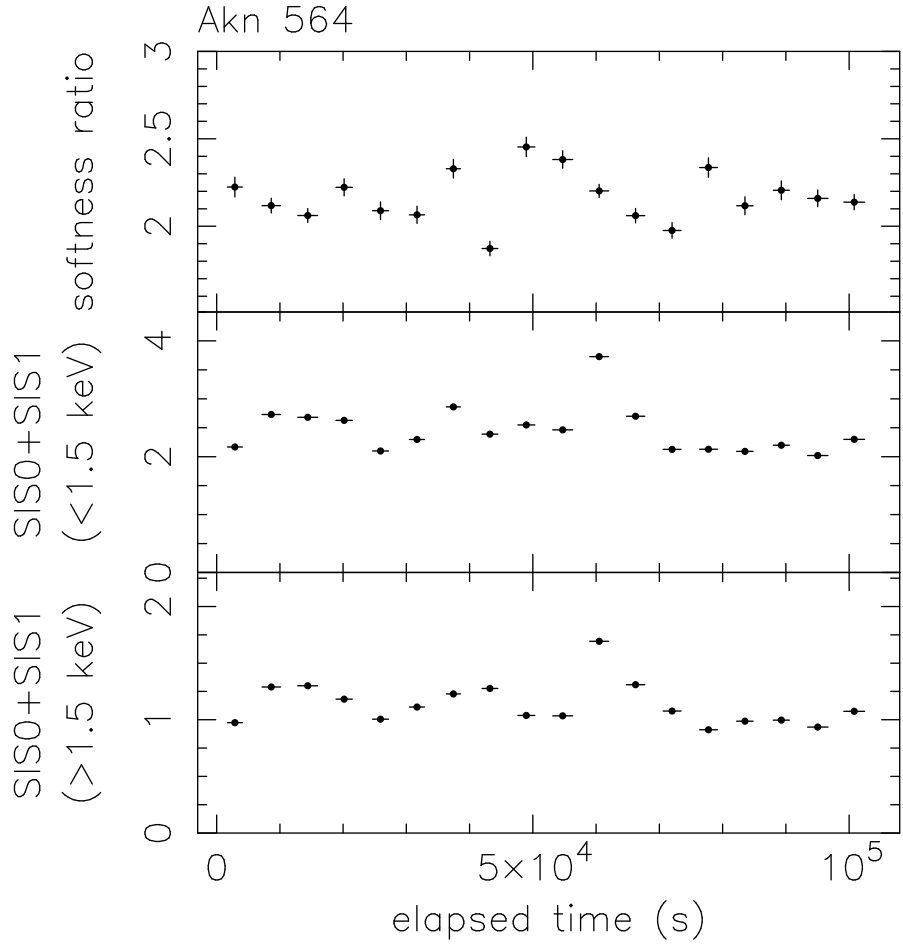}
\caption{Light curves and softness ratio from Akn 564.  The softness
is defined as the ratio of the flux below 1.5~keV to the flux above
1.5~keV.  The softness ratio variations are clearly detected and they
are uncorrelated with the flux.}
\end{figure}

\newpage

\renewcommand{\thefigure}{\arabic{figure}}
\setcounter{figure}{2}

\begin{figure}[t]
\vbox to3.5in{\rule{0pt}{3.5in}}
\includegraphics{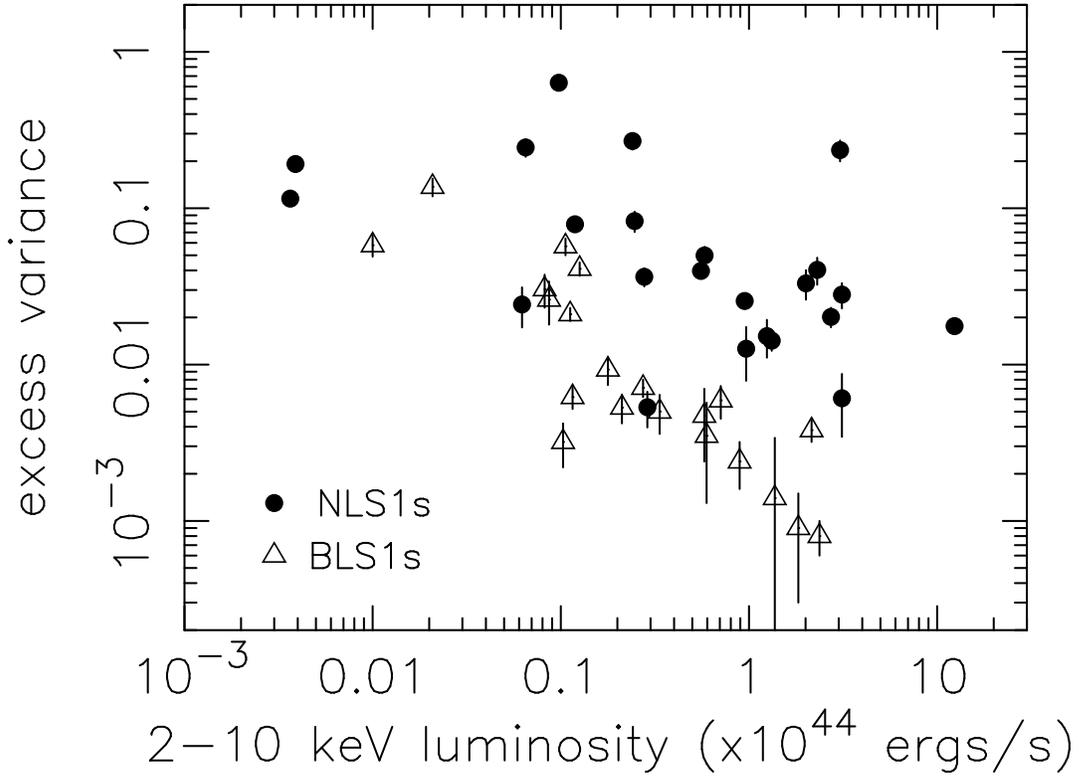}
\caption{The excess variance from {\it ASCA} observations of NLS1s
analyzed in this paper is consistently higher than that from Seyfert
1s with broader optical lines taken from Nandra et al.\ 1997a at a
particular 2--10 keV luminosity. Note that both observations of
NGC~4051 are plotted.  Mrk~507 and Kaz~163, which have low or
undetectable excess variance are not shown.}
\end{figure}

\clearpage

\begin{figure}[t]
\vbox to3.5in{\rule{0pt}{3.5in}}
\includegraphics{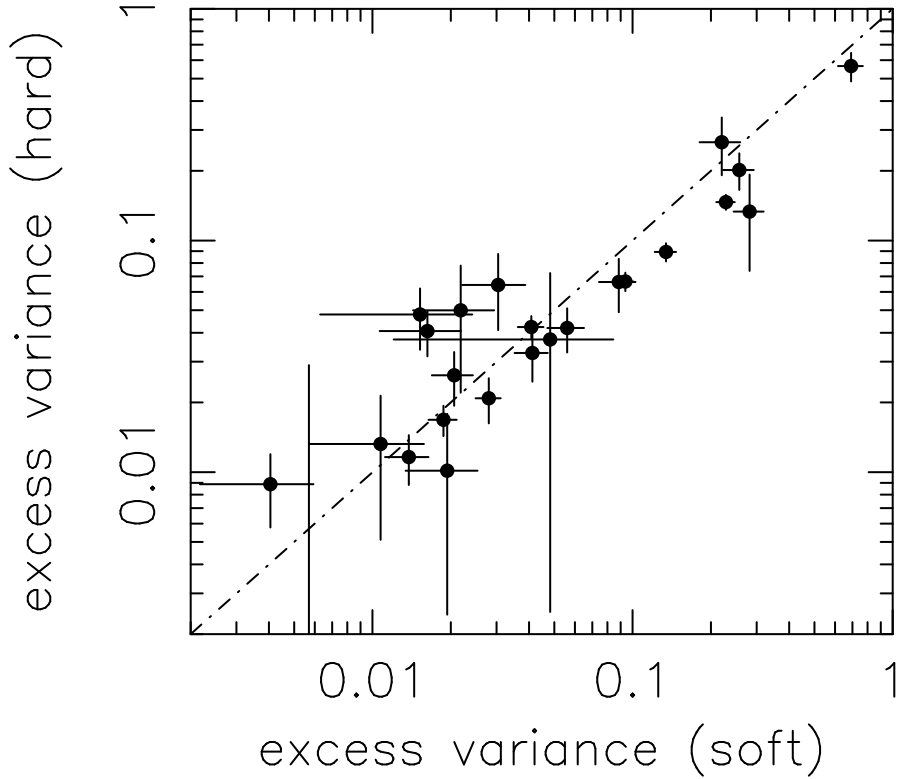}
\caption{The soft ($<1.5\rm\,keV$)
band excess variance versus the hard ($>1.5\rm\,keV$) band excess
variance.  In general, the excess variance in the soft band is slight
larger than that in the hard band.  This possibly indicates that the
spectrum becomes softer when it the object is brighter; however,
detailed analysis beyond the slope of this paper would be necessary to
confirm this.}
\end{figure}

\clearpage

\begin{figure}[t]
\vbox to5.0in{\rule{0pt}{5.0in}}
\includegraphics{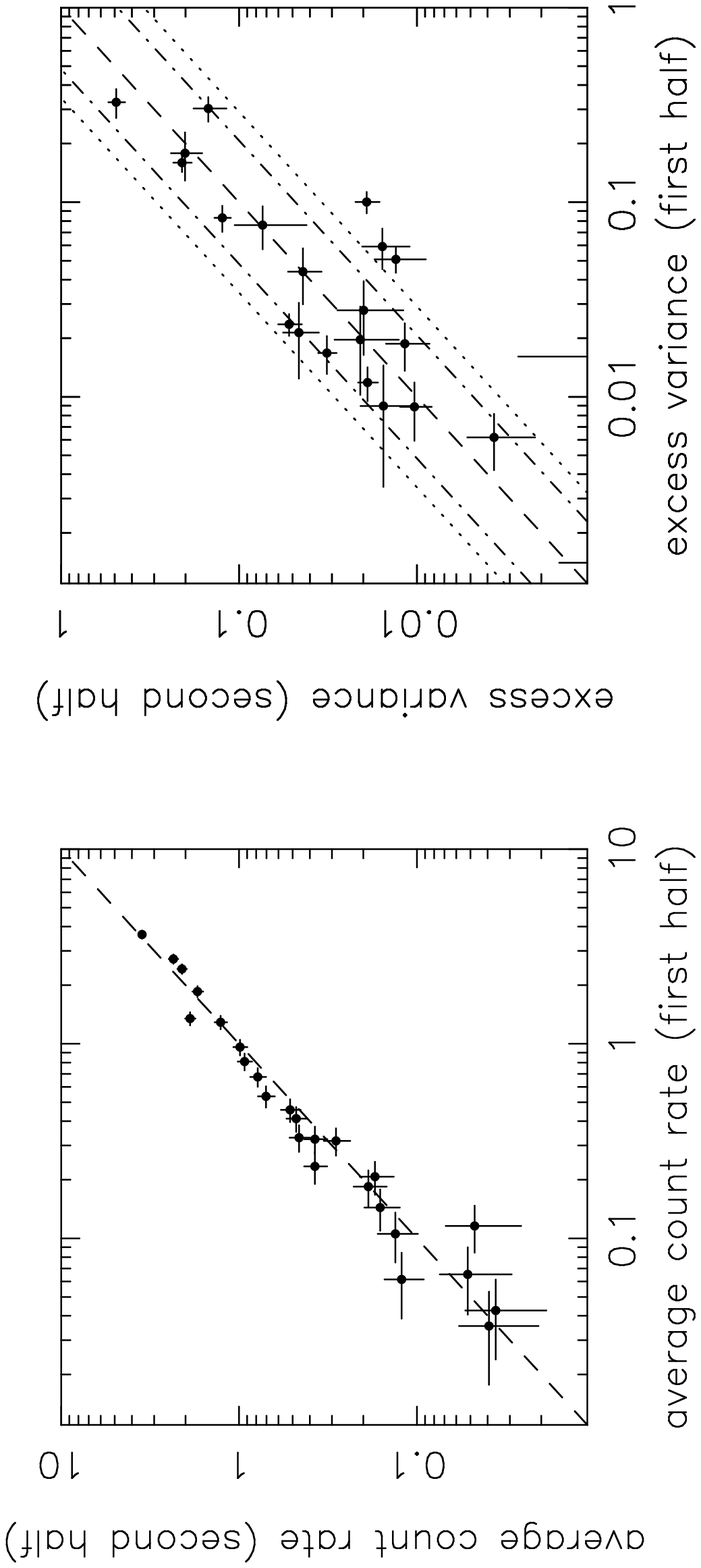}
\caption{Stationarity is tested by computing the average and excess
variance for the first and second halves (x and y axes) of the time
series.  The averages are nearly the same.  The excess variances
appear to differ, based on the error bars.  However, the error bars
reflect only measurement errors but do not consider the differences
expected due to the $1/f$ nature of the variability.  These are
explored using simulations (see text).  The dot-dash and dotted lines
shows the 68-percentile expected variance for $\alpha$=1.5 and 2.0
respectively.  Nonstationarity is indicated in I~Zw~1, Mrk~766, and
PG~1244+026, which all show a step or transition to a higher state in
their light curves.  The remaining objects are consistent with no
change in power spectrum parameters on these time scales.}
\end{figure}

\newpage

\begin{figure}[t]
\vbox to5.0in{\rule{0pt}{5.0in}}
\includegraphics{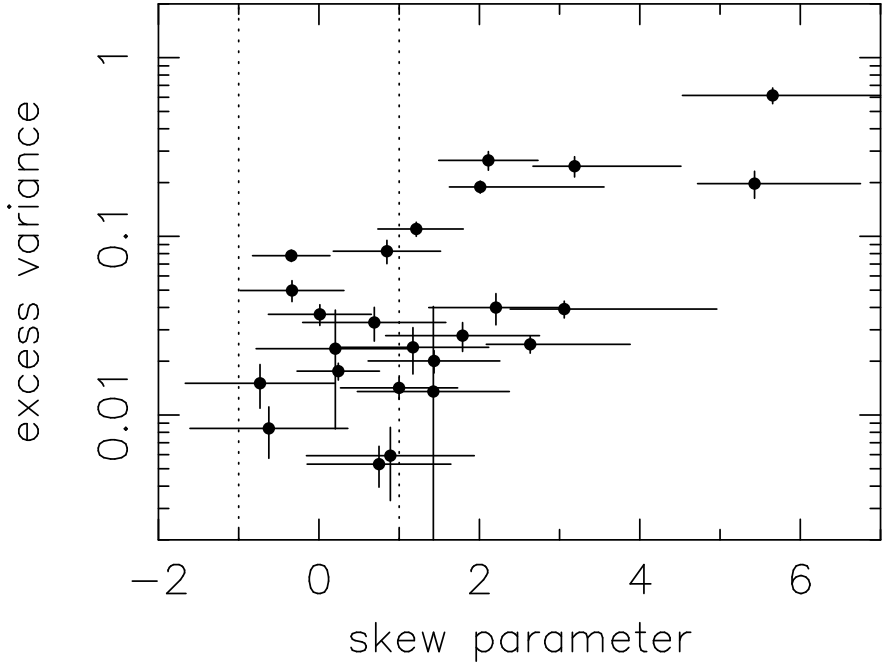}
\caption{The skew parameter versus the excess variance for the
narrow-line Seyfert 1 galaxies.  The skew parameter gives the number
of sigma significance from the null hypothesis of a nonskewed (i.e.
Gaussian) light curve, after the accidental skew due to the long term
correlations in the light curves has been accounted for using
simulations.  Since the skew parameter gives the number of sigma
significance, dotted lines are drawn at skew $=\pm 1$ to delineate
objects which have light curves in which the flux distributions are
not significantly skewed.}
\end{figure}

\newpage

\begin{figure}[t]
\vbox to5.0in{\rule{0pt}{5.0in}}
\includegraphics{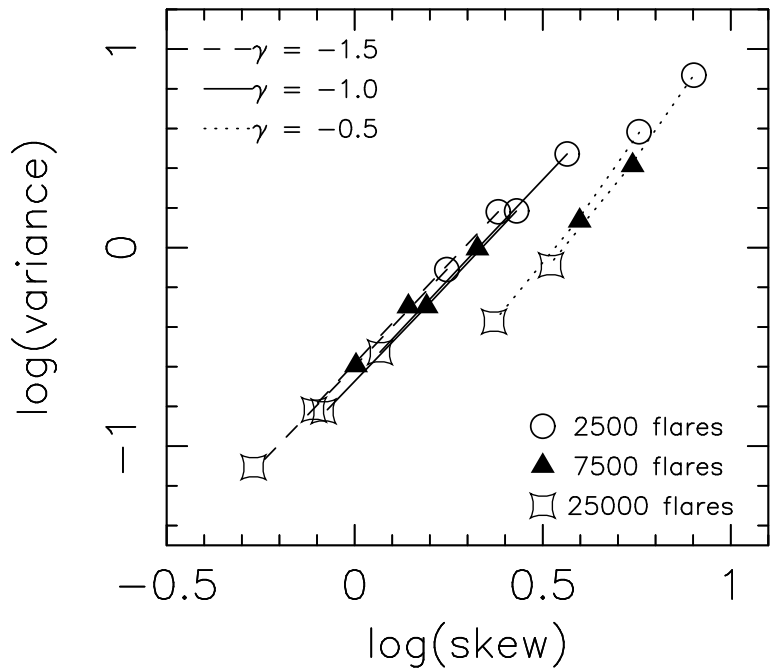}
\caption{The results from simulations from an adhoc variability model
constructed of a sum of flares (see text for a description of the
model).  For each value of $\gamma$, the results for $Nrat$=5 and
$Nrat=10$ are shown by the lower and upper lines, respectively.}
\end{figure}
\newpage

\begin{figure}[t]
\vbox to6.0in{\rule{0pt}{6.0in}}
\includegraphics{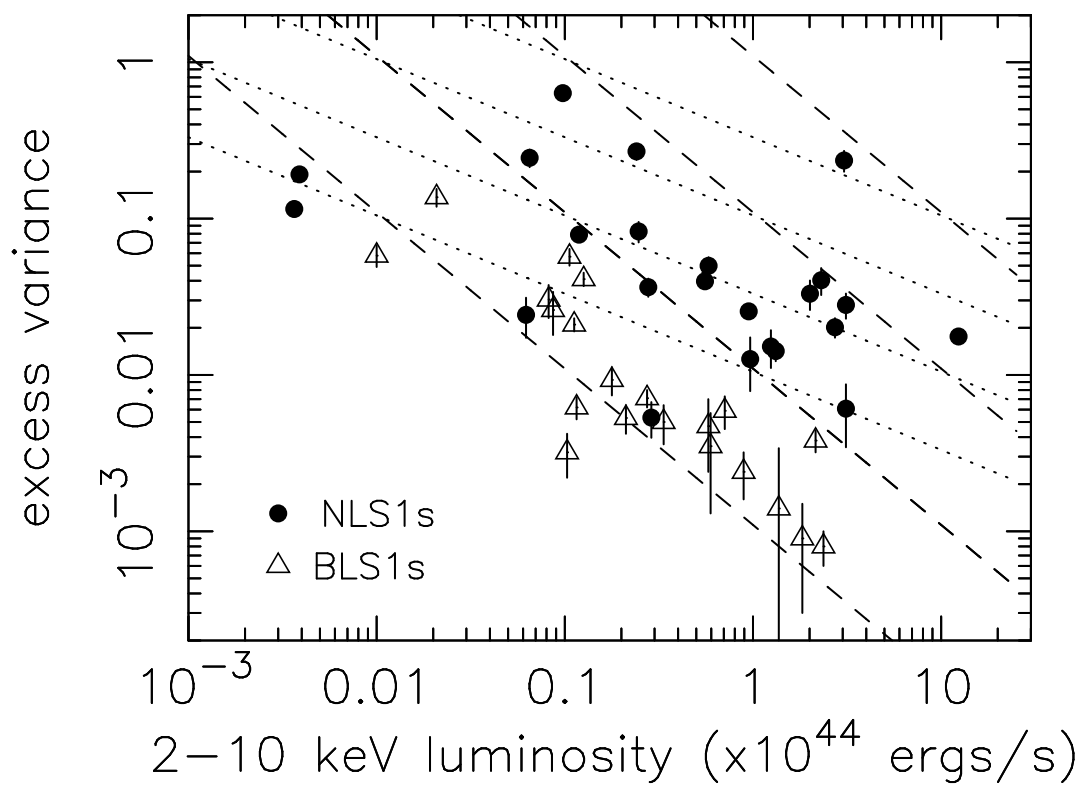}
\caption{Excess variance versus 2--10 keV luminosity, with lines
corresponding to either constant efficiency $\eta$ or constant
specific accretion rate $\dot m$ superimposed.  Dotted and dashed
lines refer to power law spectral indices 1.5 and 2.  The lines are
separated by 1 decade of the variable parameter. }
\end{figure}

\newpage

\begin{figure}[t]
\vbox to6.0in{\rule{0pt}{6.0in}}
\includegraphics{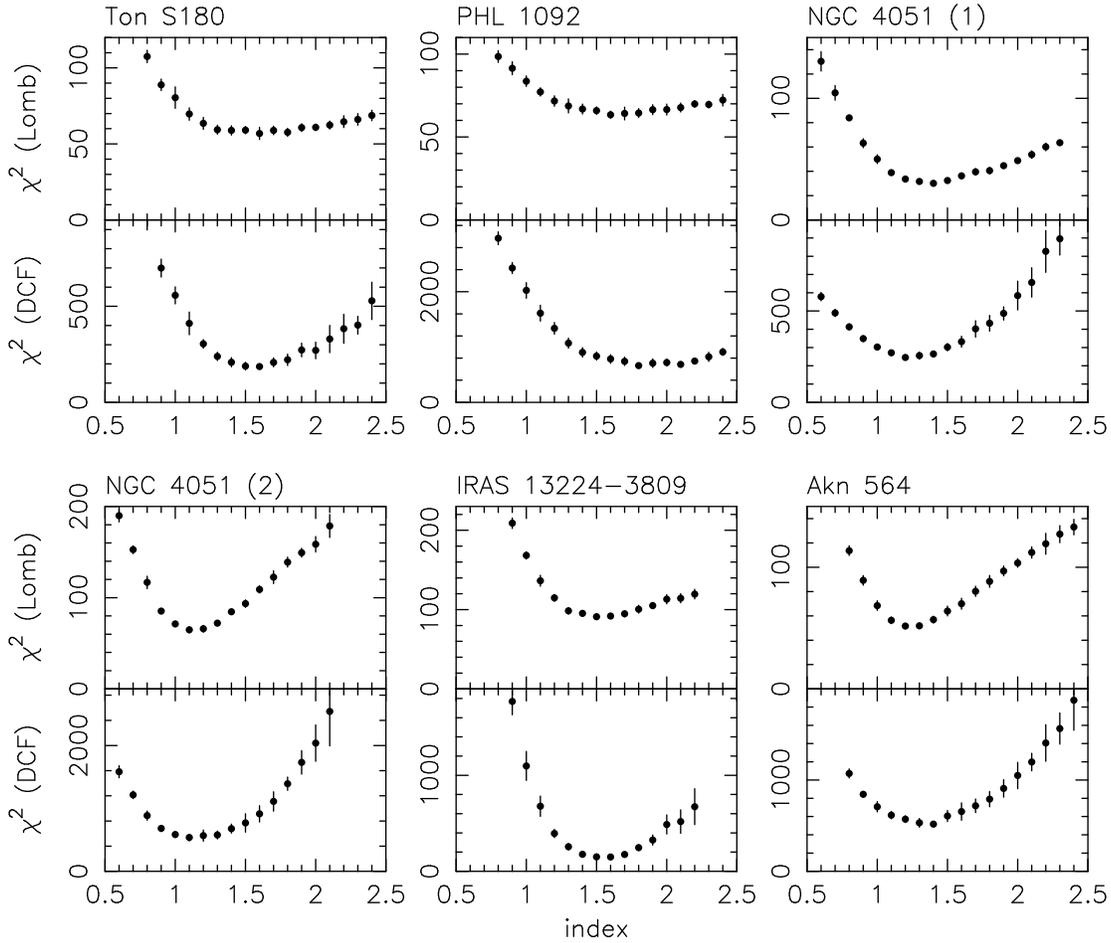}
\caption{Investigation of consistency of the variability power law
index among NLS1s.  For each object and index, one thousand simulated
light curves were made.  These were compared with the logarithmically
rebinned Lomb-Scargle periodogram (upper panel) and the discrete
correlation function (DCF; lower panel).  The mean and standard
deviations of the $\chi^2$ from 10 batches of 100 are displayed.  The
results show that minimum $\chi^2$ are roughly consistent from object
to object.}
\end{figure}

\end{document}